\documentclass[ 10pt,
 amsmath,amssymb,
 aps, physrev,
]{revtex4-2}

\usepackage{bm}
\usepackage{lipsum} 
\usepackage{xcolor}
\usepackage{amsbsy}
\usepackage{tikz}
\usepackage{float}
\usepackage{placeins}
\usepackage{footnote}
\usepackage{soul}
\usepackage{multirow}

\usepackage{eqnarray, mathrsfs}
\usepackage{subfigure}
\usepackage{colortbl}
\usepackage{xcolor}
\usepackage{footnote}
\usepackage[flushleft]{threeparttable}
\usepackage{graphicx}
\usepackage{dcolumn}
\usepackage{bm}
\usepackage[colorlinks, linkcolor=black,citecolor=blue,urlcolor=blue]{hyperref}

\begin{document}

\title{Fitness inference tested by in silico population genetics}

\author{Hong-Li Zeng}
\email{hlzeng@njupt.edu.cn}
\author{Yu-Han Huang}
 \affiliation{School of Science, Nanjing University of Posts and Telecommunications, National Laboratory of Solid State Microstructures, Nanjing University, Nanjing 210093, CHINA;\\ Key Laboratory of Radio and Micro-Nano Electronics of Jiangsu Province, Nanjing, 210023, CHINA}
\author{John Barton}%
\email{jpbarton@pitt.edu}
\affiliation{%
Department of Computational \& Systems Biology, University of Pittsburgh School of Medicine, USA
}%
\author{Erik Aurell}  
\email{eaurell@kth.se}
\affiliation{
 Department of Computational Science and Technology, AlbaNova University Center, SE-106 91 Stockholm, SWEDEN}

\date{\today}

\begin{abstract}
We consider populations evolving according to natural selection, mutation, and recombination, and assume that the genomes of all or a representative selection of individuals are known. We pose the problem if it is possible to infer fitness parameters and genotype fitness order from such data. We tested this hypothesis in simulated populations. We delineate parameter ranges where this is possible and other ranges where it is not.
Our work provides a framework for determining when fitness inference is feasible from population-wide, whole-genome, time-stratified data and highlights settings where it is not. We give a brief survey of biological model organisms and human pathogens that fit into this framework.
\begin{description}
\item[keywords]
Evolution $|$ Fitness inference $|$ Marginal
path likelihood method $|$ Transient quasi-linkage equilibrium method
\end{description}
\end{abstract}
                        
\maketitle

\section{\label{sec:level1}Introduction\protect}

Time-series genetic data, where DNA or allele frequency samples are collected from the same population at multiple points in time, offer a unique window into evolution in action. Unlike cross-sectional data collected at a single point in time, temporal samples enable the observation of changes in genetic composition as they occur. Such data also provides an excellent opportunity to reveal the forces of natural selection that drive evolution.

The ability to infer selection from temporal data could have broad biological applications. Temporal data is available to characterize evolution in microbes (e.g., experimental evolution of bacteria or viruses), humans (ancient DNA through time), and other species. For example, ancient DNA data from different epochs can reveal allele frequency shifts that signal past selective sweeps, such as the rise of lactase persistence in Europe \cite{mathieson2015genome}. Similarly, microbial evolution experiments or viral samples from successive outbreaks can show real-time selection of beneficial mutations.

However, inferring selection from temporal data is technically challenging. This task requires disentangling deterministic frequency changes caused by selection from those resulting from mutations, recombination, or random fluctuations due to genetic drift. Recent years have seen the development of a variety of methods to quantify selection in temporal data.

One popular class of methods builds explicit likelihood models for allele frequency trajectories under selection. These models are often based upon the Wright-Fisher model, a foundational model in population genetics that incorporates natural selection and other important evolutionary effects (e.g., genetic drift due to finite population size) \cite{ewens2004mathematical}. Due to technical constraints, most methods have focused on single loci \cite{bollback2008estimation, lacerda2014population, malaspinas2012estimating, mathieson2013estimating, feder2014identifying, steinrucken2014novel, ferrer2016approximate, schraiber2016bayesian, paris2019inference}. Mathematical approximations have also frequently been used to simplify analyses. For example, early work by Bollback and collaborators considered the diffusion limit of the Wright-Fisher model \cite{bollback2008estimation}, which assumes that selection is weak and the population size is large. Subsequently, other groups have developed various approximate models that aim to balance computational complexity with inference accuracy and power \cite{steinrucken2014novel, malaspinas2012estimating, lacerda2014population, paris2019inference}.

Interactions between alleles at different loci can also influence their dynamics, which cannot readily be captured in single-locus models. Examples include genetic hitchhiking, where a neutral or modestly deleterious mutation rises to high frequencies thanks to its appearance on an advantageous genetic background \cite{smith1974hitch}, and clonal interference, where multiple beneficial alleles compete \cite{muller1932some, gerrish1998fate}. Accordingly, several groups have considered the more computationally challenging task of jointly estimating selection across multiple loci \cite{illingworth2011distinguishing, illingworth2012method, foll2014influenza, terhorst2015multi, tataru2017inference}. Like the single locus methods above, these approaches also often incorporate approximations of the Wright-Fisher model to reduce computational costs. Nonetheless, these approaches are often difficult to apply to data sets that feature genetic variation at many loci \cite{Sohail2021}, which is commonly observed in viral populations.

In this work we carry out a competitive evaluation of two fitness inference schemes. The first is the marginal path likelihood (MPL) method, which is designed for data on finite populations with genetic drift and to optimally use time-ordering \cite{Sohail2021}. The second is the transient quasi-linkage equilibrium method (tQLE) which is designed for fitness landscapes with epistatic components where the quasi-linkage equilibrium phase can be expected to be present in the population \cite{Zeng_2025}. The simplest version of MPL does not include epistatic fitness (though this can be accommodated in an extended model \cite{sohail2022inferring, shimagaki2025efficient}), while tQLE deals with infinite populations where genetic drift plays no role; the two methods start from complementary assumptions on what is important to make inference feasible with the amount of data that could typically be expected to be available. 
We here show that both methods work well in inferring fitness in appropriate parameter ranges, particularly regarding the fitness order of sequences observed in some time interval.
We also show that in wide parameter ranges the two methods basically agree in their predictions, the differences in the underlying principles notwithstanding.  

In earlier work \cite{Zeng_2025} we compared the predictions of both methods on time-ordered whole-genome SARS-CoV-2 data collected in the GISAID repository\cite{GISAID}. We then found that contrary to our expectations the two methods often agreed. Logically, this could have been because both methods basically agree with ground truth, or because both methods give wrong predictions, but which happen to agree between themselves. When applied to real sequence data, and in the absence other independent estimates on fitness, there is no way to distinguish the two possibilities, although the first appears more plausible. One main motivation of the present work was hence to rule out the second possibility in a setting when ground truth is available, that is in simulations. A further objective was to map out performance in different parameter ranges going beyond the ones appropriate for SARS-CoV-2 and the COVID-19 pandemic. We are hence able to confirm that an evaluation criterion by fitness sequence order leads to better agreement both between the methods and with underlying ground truth, and in particular as to predictions on top-ranked sequences. We note that such criteria focusing on top ranked items has long been favored in the field of Direct Coupling Analysis (DCA) \cite{Weigt2009pnas, Cocco2018}, which has computational similarities to part of the tQLE procedure.

The natural evolutionary forces of population genetics are natural selection, mutation, recombination, and genetic drift \cite{Fisher-book, Blythe2007}. Selection is a fundamental driver of evolution, referring to an individual's relative probability for successful reproduction. 
In this work, we express individual genotypes through binary variables $\bm g =\{s_1,...,s_L\}$, where the indices $i$ represent sites (loci) in the genetic sequence and the $s_i\in \{-1, 1\}$ describe the state of the sequence at each site (i.e., wild-type or mutant). 
This is equivalent to assuming that all variable loci are bi-allelic (no multi-allelic loci).
For simplicity, we further assume that fitness is a quadratic function of the genetic sequence, 
\begin{equation}
    F(\bm g) = f_0 + \sum_if_is_i + \sum_{ij}f_{ij}s_is_j,
    \label{eq:fitness_func}
\end{equation}
with $f_i$ representing the additive effects or selection coefficients from $L$ single loci and $f_{ij}$ the epistatic effects from pairs;
$f_0$ is an overall shift which can be used to ensure that \textit{e.g.} the fitness of all genomes are non-negative.
The MPL method that we consider relies only on the additive fitness coefficients $f_i$, and non-zero epistatic coefficients 
$f_{ij}$ act as confounding factors for which the method can be stress tested. 
The tQLE method on the other hand uses both $f_i$ and $f_{ij}$, and can be used to infer $f_{ij}$. 
Higher-order epistasis is not considered in this work. 
In both cases our main evaluation criterion will be the order of sequences as to fitness. In MPL the inferred fitness order of sequences is determined by the inferred additive fitness parameters $f_i$, while in tQLE the inferred fitness  order of sequences is determined by both inferred additive fitness parameters $f_i$ and inferred epistatic fitness parameters $f_{ij}$.

Given a set of time-stamped genomes, either from data, or, as here, from simulations,
different approaches can be applied to infer fitness parameters.
In this work we use two methods for fitness inference, specifically for the selection coefficients. 

The marginal path likelihood approach (MPL) \cite{sohail2022inferring} is based on the evolution of nucleotide or amino acid frequencies in the Kimura diffusion approximation \cite{Kimura1956, Kimura1964}. The starting point is the joint probability $P\left(\{m\}^{(1)}, \{m\}^{(2)}, ..., \{m\}^{(L)}\right)$ with $m_a^{(i)}$ the normalized frequency of allele $a$ at locus $i$. This probability
satisfies a Fokker-Planck equation (Kolmogorov forward equation), for which a path-wise counter-part is a generalized Langevin equation. Maximizing the probability of an observed temporal sequence
$m_a^{(i)}(t)$ over the parameters of this generalized Langevin equation is the key step in the MPL procedure.

The transient Quasi-Linkage Equilibrium (tQLE) approach assumes that the state of the nucleotide during the evolutionary process follows the Gibbs-Boltzmann distribution transiently \cite{Zeng_2025}. 
\begin{equation}
      P(\bm{g})= \frac{1}{Z}
        \exp\big(\sum_{i,a} h_{i,t}(a)\bm{1}_{s_i,a} +\sum_{i<j,ab} J_{ij,t}(a,b) \bm{1}_{s_i,a}\mathbf{1}_{s_j,b}\big)
         \label{eq:Ising-Potts}
\end{equation} 
The time $t$ in \eqref{eq:Ising-Potts} indicates that the Ising parameters $h_{i,t}$ and $J_{ij,t}$ are not necessarily constant, and \eqref{eq:Ising-Potts} is not based on the assumptions that lead to thermal equilibrium with constant parameters. The relation between the Ising parameters ($h_i$, $J_{ij}$) and the fitness parameters ($f_i$,  $f_{ij}$) is the key point of the QLE theory. 

In the COVID-19 pandemic, tens of millions of full-length high-quality genomes of the etiological agent were collected and in many cases uploaded to the GISAID repository\cite{GISAID}.
One of the more extensively sequenced countries was the UK, with a maximum of about 40,000 unique SARS-CoV-2 sequences uploaded per week in December 2021\cite{Zeng_2025}.   
The prevalence of SARS-CoV-2 infections at any given time has been estimated to be on the order of one percent, \textit{i.e.}, 
in the total human population of
a hundred million individuals. Assuming a $1\%$ infection rate in the UK at peak, the sampling ratio reached $10\%$. In most other areas of the world (and in the UK outside the time window of late 2021) it was much less. Nevertheless, compared to earlier epidemics, this was unprecedented in detail, and is not yet rivaled for another virus.

More recently, the GISAID-led surveillance efforts for influenza resulted at its peak a comparable number of sequences (tens of thousands per week at the end of 2024, though globally). 
The NCBI Pathogen Detection integrates bacterial and fungal genomic sequences \cite{NCBI-Pathogens} of which there are often hundreds of new entrants per day, recently for \textit{Salmonella}.

Beyond the genomic surveillance of pathogens, there are other sources of genomic time series data that could yield important insights into evolution.
One source is ancient DNA \cite{paabo2004genetic, orlando2021ancient}, isolated from preserved biological tissues. The oldest such samples have been obtained from more than a million years before the present day \cite{kjaer20222}, opening the possibility for studying evolution over vast time scales.
Data from experimental evolution can also provide insights into microbial adaptation in artificial environments. The \textit{E. coli} long-term evolution experiment\cite{lenski1991long}, originated by Richard Lenski, is one of the most famous examples. But a new generation of experimental approaches has also been developed to probe evolution with increasing levels of detail and across a wide variety of conditions \cite{ascensao2025experimental}.

The paper is organized as follows. 
In Section~\ref{sec:materials} we present 
the synthetic data and the methods we used for 
further analysis.
In Section~\ref{sec:results}
we present the results of we study and 
in Section~\ref{sec:discussion}
we discuss implications and future work.
The estimated fitness and the inferred rank order of the top-performing sequences Oby MPL and tQLE for the epistatic case 

\section{Materials and Methods}
\label{sec:materials}
The natural evolutionary forces of population genetics are selection, mutation, recombination, and genetic drift \cite{Fisher-book, Blythe2007}. Selection is a fundamental driver of evolution, referring to the hereditary probability of an individual during the reproduction process. This is closely related to the fitness function of given genotypes, which can be described by Eq.~\eqref{eq:fitness_func}.
For simplicity we have here not included the constant offset term $f_0$, which cannot be inferred in the models and with the data considered, as the reproductive success of an individual depends only on relative fitness. 
Only quadratic contributions to epistatic fitness are considered in this work.
Simulations of the evolutionary genetic population can be performed through the toolbox \mbox{FFPopSim} \cite{FFPopSim}. 
With the synthetic genomic data, two inference methods for the additive effects are compared. One is the transient Quasi-Linkage Equilibrium (tQLE) approach, which assumes the state of the nucleotide during the evolutionary process follows the Gibbs-Boltzmann distribution transiently \cite{Zeng_2025}. The other is the Marginal Path Likelihood (MPL) approach. The corresponding procedures of these two methods are explained as follows.

\subsection{Additive and epistatic fitness inference using tQLE}

As stated in ref.~\cite{Zeng_2025}, QLE is a dynamic theory where parameters $(\{h\},\{J\})$ in general change in time. We here introduce the derived abbreviation tQLE to emphasize that we use the formulas for inference on such in principle (and generally in practice) time-changing data.

The equation for $(\{J\})$ is of the relaxation type, and in the theory of \cite{NeherShraiman2011} 
\begin{equation}
    \dot{J}_{ij}(a,b) = f^{(2)}_{ij}(a,b) - r c_{ij} J_{ij}(a,b)
\end{equation}
For large enough $r$ the Potts parameters will hence relax to a stable fixed point, \textit{i.e.} to $J^*_{ij}(a,b) = \frac{f^{(2)}_{ij}(a,b)}{r c_{ij}}$, which allows to \textit{infer} epistatic fitness parameters from Potts parameters computed from the data through the formula
\begin{equation}
    \label{eq:f-2-inference}
    f^{(2),*}_{ij}(a,b) = r c_{ij} J^*_{ij}(a,b)
\end{equation}
This relation was derived in \cite{NeherShraiman2011}, and tested (in the stationary state) in \cite{Zeng2020}. In \cite{Zeng2021JStat} \eqref{eq:f-2-inference} is extended in wider parameter ranges (\textit{e.g.} large mutation rates) to  
\begin{equation}
\label{eq:f-2-inference-GC}
    f^{(2),*}_{ij}(a,b) = (r c_{ij}+4\mu) J^*_{ij}(a,b)
\end{equation}
and 
\begin{equation}
\label{eq:f-2-inference-corr}
    f^{(2),*}_{ij}(a,b) = \frac{(r c_{ij}+4\mu) \chi_{ij} (a,b)}{(1-\chi_i^2)(1-\chi_j^2)}
\end{equation}
with a Gaussian closure technique. 
As discussed in \cite{Zeng2022}
since \eqref{eq:f-2-inference} only
relates pair-wise quantities, it can also
work when the single-nucleotide frequencies
change. This can, for instance, be the case 
of additive fitness changes in time, say
by a change in the fitness landscape of which one example could be the introduction of widespread vaccination against SARS-CoV-2 in the COVID-19 pandemic.

The equation for $(\{h\})$ is on the other hand not of the relaxation type
(\cite{NeherShraiman2011}, Eq.~24)
\begin{equation}
    \dot{h}_{i}(a) = f^{(1)}_{i}(a) + r \sum_{j,b} c_{ij} 
    J_{ij}(a,b) m_j(b)
\end{equation}
where $m_j(b)=\sum_{\mathbf{x}} P(\mathbf{x}) \mathbf{1}_{x_j,b}$ is the frequency of allele $b$ at locus $j$.
Combining \eqref{eq:f-2-inference} and inferred values of $\{h\}$ at two consecutive time intervals lead to the inference formula
\begin{eqnarray}
    f^{(1),*}_{i}(a, \Gamma) &=& \frac{1}{\Delta t}\left[h_{i}(a, \Gamma + \Delta t) -
    h_{i}(a,\Gamma)\right] 
    \nonumber \\
    \label{eq:f-1-inference-QLE}
    &&
    - \sum_{j,b}  f^{(2),*}_{ij}(a,b, \Gamma)\, m_j(b,\Gamma)\,,
\end{eqnarray}
with $\Gamma$ indicating the discrete time. 

The direct effects $h_i$ acting on locus $i$ is obtained through the naive mean-field (nMF) approximation 
\begin{equation}
\label{eq:nMF_mi}
    h_i^*(\Gamma) = \tanh^{-1}m_i(\Gamma)-\sum_jJ_{ij}m_j\left(\Gamma\right)
\end{equation}
The direct couplings  $J_{ij}$ between locus $i$ and $j$ are computed using all replicates accumulated up to the current generation during genomic evolution.

For the case with no epistasis, the nMF approximation of $m_i$ takes the first term of \eqref{eq:nMF_mi}, which leads to the inference formula for the additive effects that takes the first term of \eqref{eq:f-1-inference-QLE}.  The corresponding fitness of an individual genomic sequence is the first additive term of \eqref{eq:fitness_func} as well.

\subsection{Additive fitness inference by MPL}

The marginal path likelihood (MPL) method \cite{Sohail2021} is based on the 
evolution of nucleotide frequencies in Kimura's diffusion approximation
\cite{Kimura1956, Kimura1964}. The starting point is thus the joint probability
$P(\{m\}^{(1)},\{m\}^{(2)},\ldots,\{m\}^{(L)})$ where $m^{(i)}_a$ is the frequency of allele $a$ on locus $i$, normalized as $\sum_a m^{(i)}_a = 1$.
In the diffusion approximation, this probability satisfies
a Fokker-Planck equation
\begin{equation}
    \label{eq:diffusion}
    \partial_t P
    = - \sum_{i,a} \frac{\partial}{\partial m^{(i)}_a}
    \left(u^{(i)}_a P\right)
    + \sum_{ij,ab} \frac{\partial^2}{\partial m^{(i)}_a\partial m^{(j)}_b}
    \left(D^{(ij)}_{ab} P\right)
\end{equation}
where the drift vector and diffusion matrix are given by
(\cite{Sohail2021}, Eq 6 and Eq S9 and following, notation aligned with the present presentation)
\begin{eqnarray}
u^{(i)}_a &=&  m^{(i)}_a (1-m^{(i)}_a)f^{(1)}_i(a) 
+ \mu\left(1- 2 m^{(i)}_a\right) 
\nonumber \\
    \label{eq:drift-vector}
&&\quad
+ \sum_{j,b} \left( 
m^{(ij)}_{ab} - m^{(i)}_a m^{(j)}_b\right) f^{(1)}_j(b) \\
    \label{eq:diffusion-matrix}
D^{(ij)}_{ab} &=& \{ \begin{array}{lr}
m^{(i)}_a m^{(i)}_b & i=j \\
m^{(ij)}_{ab} - m^{(i)}_b m^{(j)}_b & i\neq j
\end{array}
\end{eqnarray}

The Fokker-Planck equation \eqref{eq:diffusion} corresponds to a multidimensional Langevin equation, for which the probability of a path sampled at discrete times can be estimated by standard arguments. Maximizing this path probability with a Gaussian prior leads to the central inference formula in MPL 

\begin{eqnarray}
&&f^{(1),*}_{i}(a) =
\sum_{j,b} 
\left[\sum_{k=1}^K \Delta t_k D^{(ij)}_{ab}(t_k)+\gamma\mathbf{1}_{ia,jb}\right]^{-1}_{ia,jb}
\nonumber \\
        \label{eq:f-1-inference-MPL}
&&
\left[m^{(j)}_b (t_K)-m^{(j)}_b (t_0) -\mu 
\sum_{k=1}^{K-1} \Delta t_k \big(1-2 m^{(j)}_b(t_k)\big)\right] ~~~
\end{eqnarray}
In above a time interval $[t_0,t_K]$ has been divided up in $K$ sampling
intervals and the allele frequencies ($m^{(i)}_a$) and drift and diffusion
terms (from \eqref{eq:drift-vector} and \eqref{eq:diffusion-matrix})
estimated for each. The sampling interval times are defined as
$\Delta t_k = t_{k+1}-t_k$. $\gamma$ is the width of the Gaussian prior,
and acts as a regularizer.

In \eqref{eq:f-1-inference-QLE} the inferred additive fitness depends on time
and is linear in the time derivative of one inferred Potts parameter 
$h_{i}(a)$. This parameter is in itself a (complicated) function of the 
single-nucleotide and pair-wise frequencies at that time. 
In \eqref{eq:f-1-inference-MPL} the inferred additive fitness
also depends on time, more specifically on a time interval, and is linear
in $m^{(i)}_a (t_K)-m^{(i)}_a (t_0)$, the change in all the single-nucleotide
frequencies over that interval. Pair-wise frequencies also enter in \eqref{eq:f-1-inference-MPL}, through the dependence of $D^{(ij)}_{ab}$
as in \eqref{eq:diffusion-matrix}.

\section{Results\protect}
\label{sec:results}

Under what conditions can multilocus methods recover genotype fitness from genetic time-series data?
To address this question, we compared two approaches that make complementary assumptions about evolutionary dynamics: the transient Quasi-Linkage Equilibrium (tQLE) framework and marginal path likelihood (MPL). 
Beyond direct comparisons between these methods, we sought to identify parameter regimes where inference is more or less difficult, and to understand the connection between fitness estimates at the level of individual alleles or epistatic interactions and genotype fitness.


We used \mbox{FFPopSim} \cite{FFPopSim} to efficiently simulate the evolution of a population of haploid individuals.
\mbox{FFPopSim} jointly models the effects of natural selection, mutation, recombination, and genetic drift.
We considered two fitness architectures: (i) \emph{additive-only}, where loci contribute independently to fitness; and (ii) \emph{additive + pairwise epistasis}, where fitness includes pairwise interaction terms, but higher-order interactions are absent.
All simulations had a population size of $N=1000$ individuals and considered a genetic sequence of length $L=25$, evolved over $T=30$ generations, with a strong recombination rate of $r=0.5$.

\begin{table*}[b!]
\centering
\caption{Summary of parameters in simulations of evolving populations in MS (main body of the paper)}
\begin{ruledtabular}
\begin{tabular}{l|ccccllll}
Symbol & $N$ & $N_{\mathrm{realizations}}$ & $L$ &
$T$ & $r$ & $\mu$ & $\sigma(f_i)$ & $\sigma(f_{ij})$
\\
Name & Population size & Replicates & Genome size &
Time & Recomb. rate & Mutation rate & Additive fitness & Epistatic fitness
\\
\hline
Figs~\ref{fig:scatter_plts_for_fi}, \ref{fig:scatter_plts_for_estimated_additive_fitness}, \ref{fig:scatter_plts_for_ranks_of_additive_fitness} &
1000,1000,1000 & 1,1,1 &
25,25,25 & 30,30,30 & 0.5,0.5,0.5 &
0.003,0.01,0.01 & 0.01,0.01,0.05 & 0.0,0.0,0.0
\\
Figs~\ref{fig:histograms_for_differnt_fis}, \ref{fig:scatter_plts_for_simulation_dataset}, \ref{fig:scatter_plts_F1_F2}  &
1000,1000,1000 & 30,30,30 &
25,25,25 & 30,30,30 & 0.5,0.5,0.5 &
0.01,0.01,0.01 & 0.005,0.05,0.1 & 0.002,0.002,0.002 
\\
\end{tabular}
\end{ruledtabular}
\footnotetext{This table summarizes and completes 
the descriptions of simulation parameters given in figure caption
to respective figure. Multiple numerical values indicate 
values in different panels, top to bottom.
In Fig~\ref{fig:scatter_plts_for_fi} and
Fig~\ref{fig:scatter_plts_for_estimated_additive_fitness} 
first mutation rate $\mu$ and then $\sigma(f_i)$, the
variability (standard deviation) of the
additive fitness parameters $F_i$ (Gaussian random numbers),
are varied, at zero epistatic fitness parameters $F_{ij}$.
In 
Fig~\ref{fig:scatter_plts_for_ranks_of_additive_fitness},
Fig~\ref{fig:histograms_for_differnt_fis} 
and Fig~\ref{fig:scatter_plts_F1_F2}
first mutation rate $\mu$ and then $\sigma(f_i)$, 
variability (standard deviation) of the
additive fitness parameters $F_i$ (Gaussian random numbers),
are varied, at non-zero epistatic fitness parameters $F_{ij}$.
The parameter $\sigma(f_{ij})$ indicates the 
variability (standard deviation) of the
epistatic fitness parameters $F_{ij}$ (Gaussian random numbers).
}
\label{tab:parameter-values}
\end{table*}

\subsection{Additive-only fitness landscapes} 

A purely additive fitness landscape provides a simpler evolutionary setting to test inference approaches.
Here, we varied the mutation rate, $\mu$, to consider both weaker ($\mu = 0.003$) and stronger ($\mu = 0.01$) forces of mutation.
In our simulations, we assumed that the fitness effects of mutations $f_i$ were distributed according to a normal distribution with mean of zero and standard deviation of $\sigma(f_i)$. 
As for the mutation rate, we considered weaker ($\sigma(f_i)=0.01$) and stronger ($\sigma(f_i) = 0.05$) effects of selection.
Given a population size of $N=1000$, both the mutation rates and typical strengths of selection are comparable to or greater than the strength of genetic drift ($\sim 1/N = 0.001$).
In this scenario, we tested the performance of each method using a single simulation replicate as input.

\begin{figure}[t!]
\centering
\includegraphics[width=.3\linewidth]{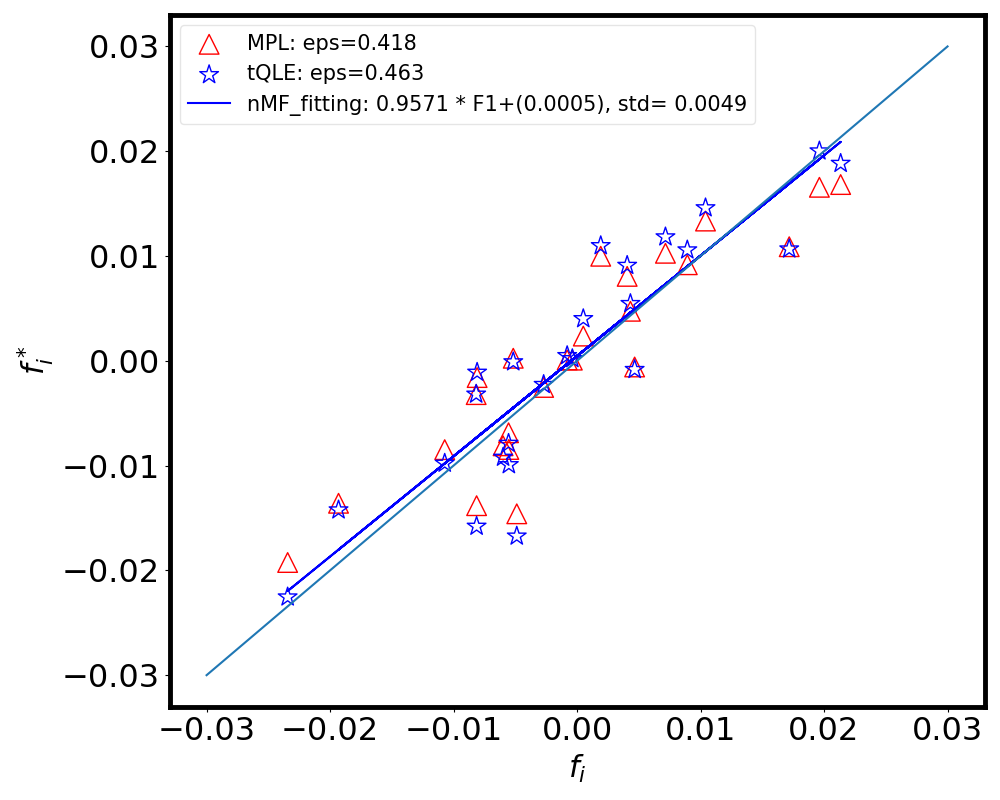}
\includegraphics[width=.3\linewidth]{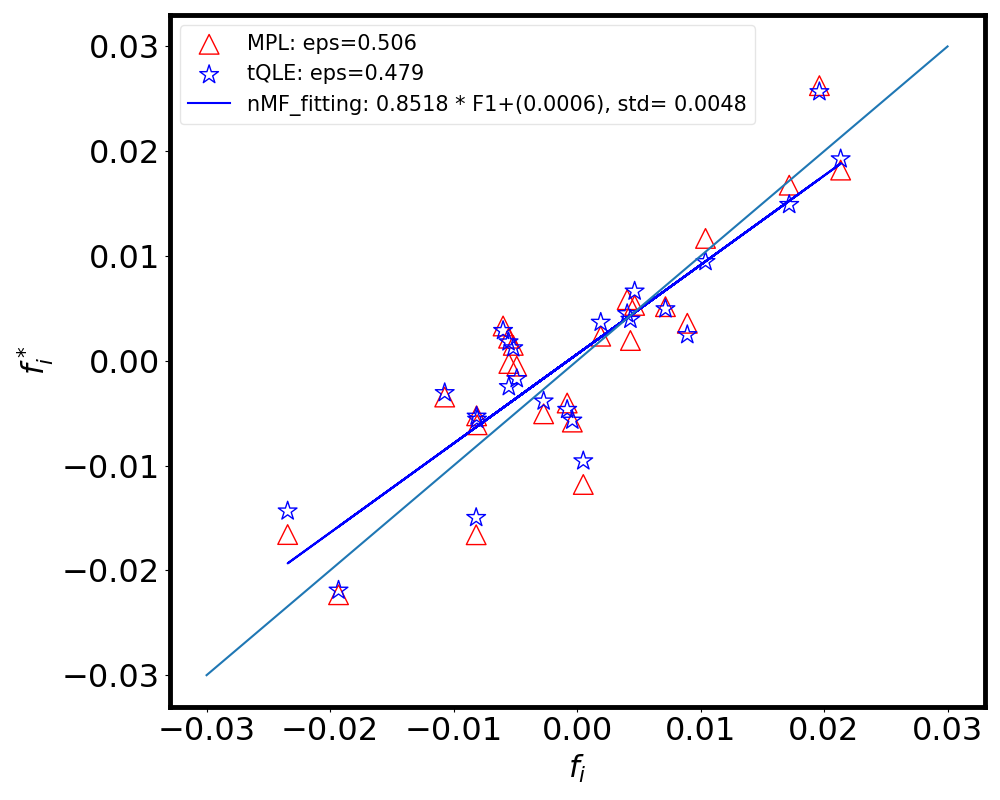}
\includegraphics[width=.3\linewidth]{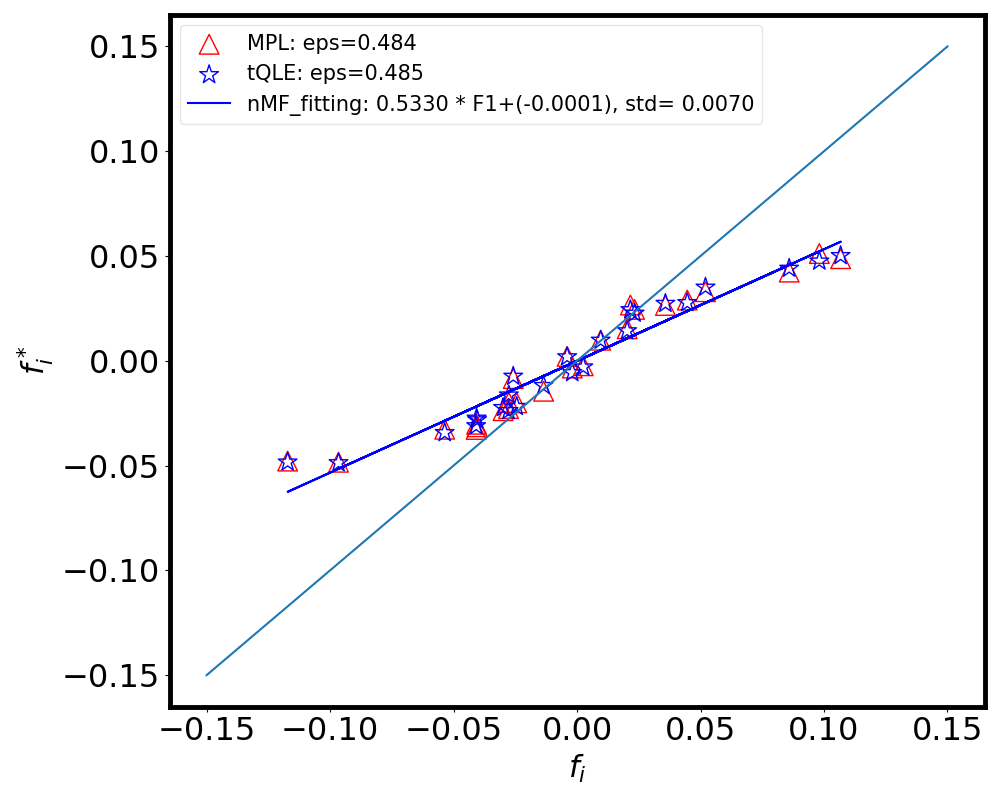}
\caption{Scatter plots for the reconstructed additive fitness $f_i^*$ versus the ground truth $f_i$ by the tQLE (blue) and MPL (red) approaches, displaying the effects 
of increasing mutation rate
(parameter $\mu$)
and increasing fluctuating additive fitness
(parameter $\sigma(f_i)$). 
Upper: $\mu = 0.003$, $\sigma(f_i) = 0.01$, middle:  $\mu = 0.01$, $\sigma(f_i) = 0.01$, bottom: $\mu = 0.01$, $\sigma(f_i) = 0.05$.
Fitness parameters $f_i$ are drawn randomly
from a Gaussian distribution with mean zero and
standard deviation $\sigma(f_i)$.
In these simulations the underlying ground truth
is only additive fitness (no epistatic fitness),
and corresponding model parameter therefore $\sigma(f_{ij})=0$.
Other parameter values as indicated in Table~\ref{tab:parameter-values}.
}
\label{fig:scatter_plts_for_fi}
\end{figure}

Overall, we found that tQLE and MPL performed similarly in these scenarios. 
Both approaches were able to accurately infer the fitness effects of individual mutations (Fig.~\ref{fig:scatter_plts_for_fi}).
In general, inference was more accurate on an absolute scale when selection and mutation were weaker.
Increasing the variance of the selection coefficients naturally led to a wider range in fitness values. These were recovered well on a relative scale, but we found that both approaches tended to underestimate the magnitude of selection for the most impactful mutations under these conditions.

\begin{figure}[h!]
\centering
\includegraphics[width=.3\linewidth]{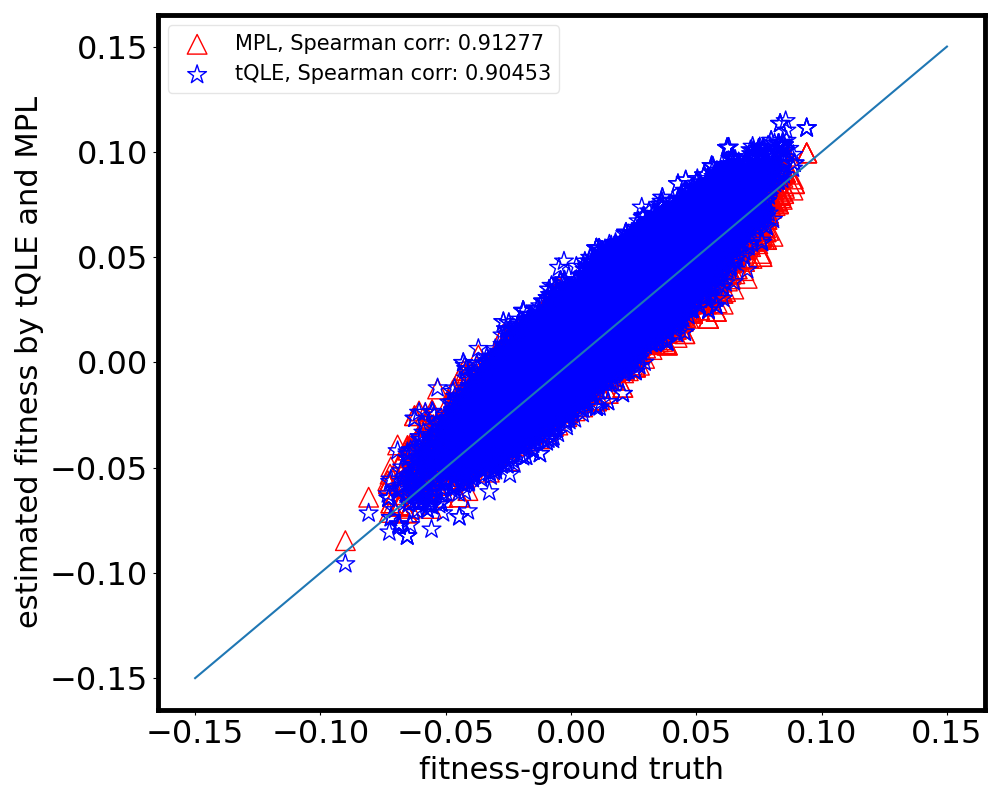}
\includegraphics[width=.3\linewidth]{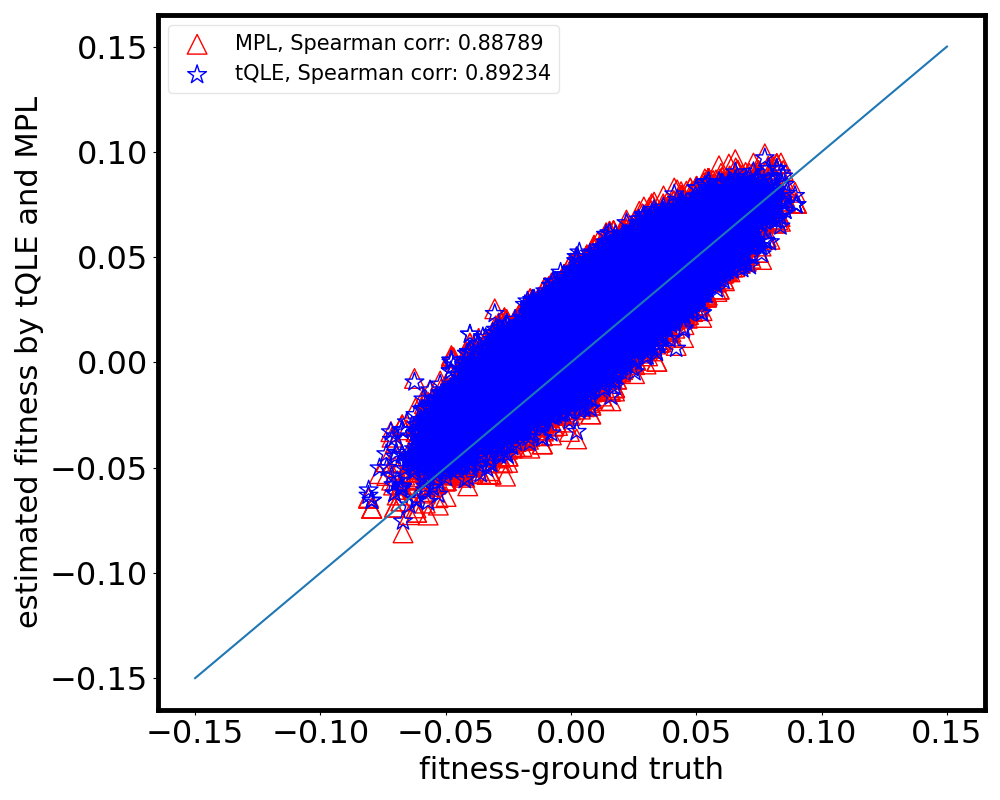}
\includegraphics[width=.3\linewidth]{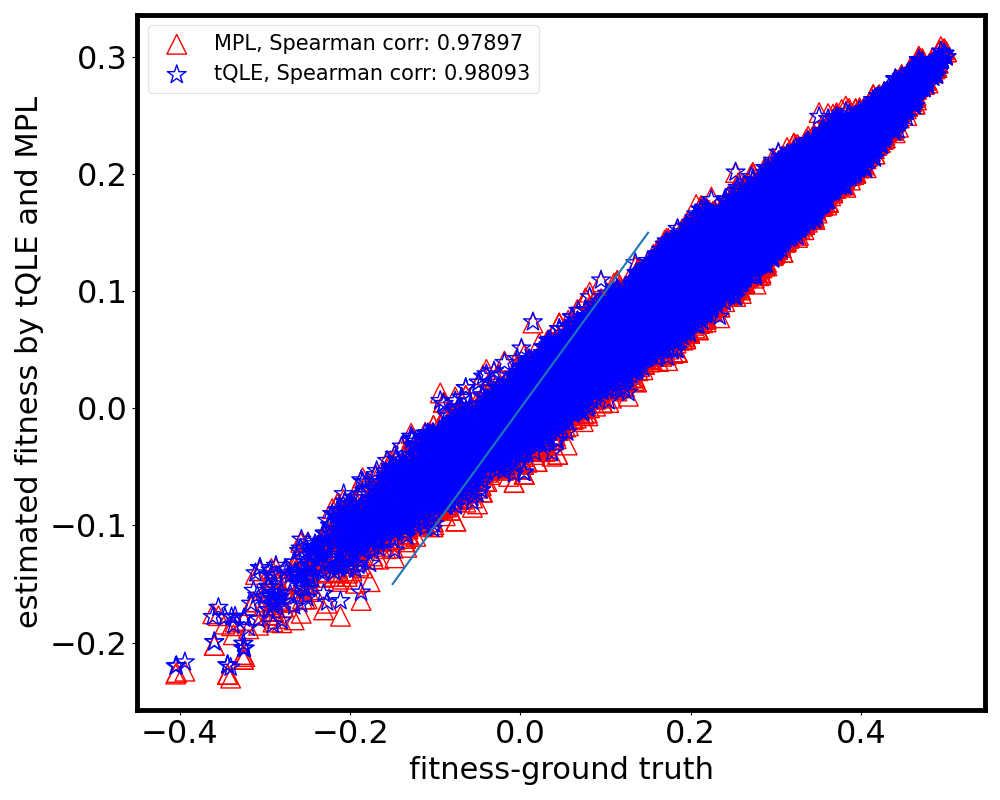}
\caption{Scatter plots for total fitness
of sequences
inferred by MPL and tQLE against ground-truth fitness values 
displaying the effects of increasing mutation rate (parameter $\mu$) and increasing fluctuating additive fitness (parameter $\sigma(f_i)$, with the same parameter choices
as in~Fig. \ref{fig:scatter_plts_for_fi}.
Red triangles represent MPL results, and blue stars represent tQLE results.
In the tests reported here only the additive fitness part of tQLE is taken into account \textit{i.e.} inferred epistatic fitness $f_{ij}^*$ is set to zero throughout, including 
as it enters in the formula for inferred 
additive fitness~\eqref{eq:f-1-inference-QLE}.
In these simulations the underlying ground truth
is only additive fitness (no epistatic fitness),
and corresponding model parameter therefore $\sigma(f_{ij})=0$.
Other parameter values as indicated in Table~\ref{tab:parameter-values}.
}
\label{fig:scatter_plts_for_estimated_additive_fitness}
\end{figure}

Similarly, both tQLE and MPL produced tightly aligned reconstructions of total genotype fitness across conditions, with little sensitivity to the mutation rate $\mu$ at fixed $\sigma(f_i)=0.01$ (Fig.~\ref{fig:scatter_plts_for_estimated_additive_fitness}). 
When the additive signal is larger ($\sigma(f_i)=0.05$), the global rank correlation with the true, underlying fitness tends to increase, but this comes with heavier deviations from the identity. 
That is, as in the recovery of individual selection coefficients, the inferred magnitude of the highest and lowest fitness values is underestimated compared to the true ones. 
These patterns are consistent with stronger signals aiding in correctly ordering fitness effects, but leading to larger errors on an absolute scale.

\begin{figure}[h!]
\centering
\includegraphics[width=.3\linewidth]{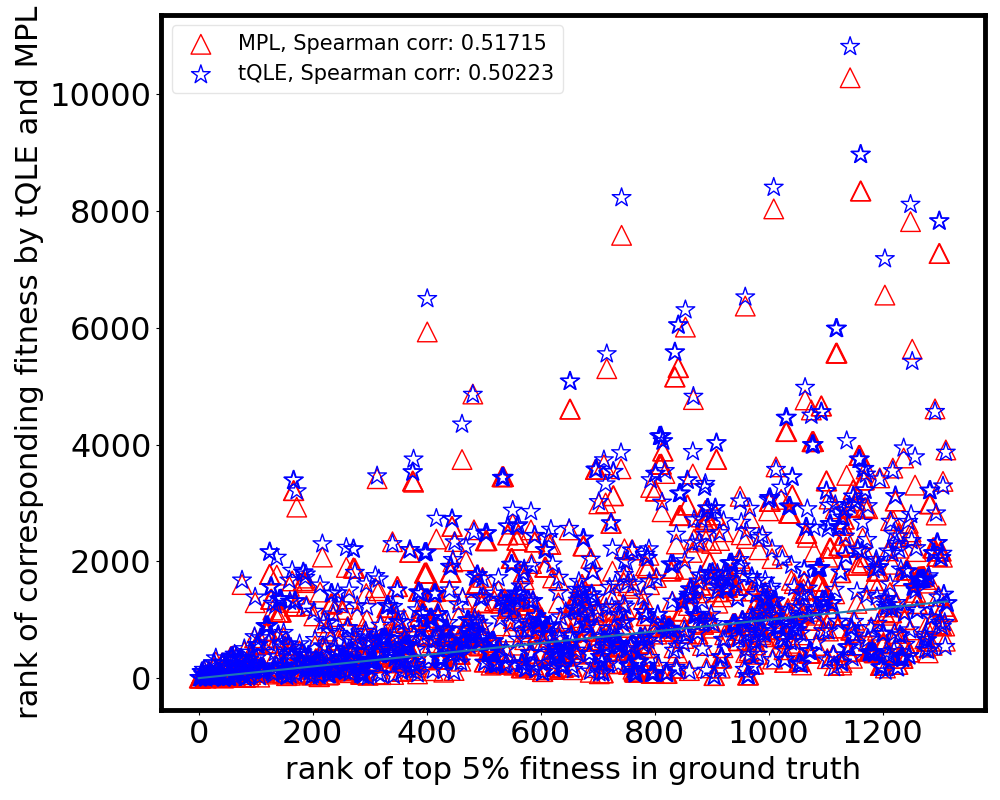}
\includegraphics[width=.3\linewidth]{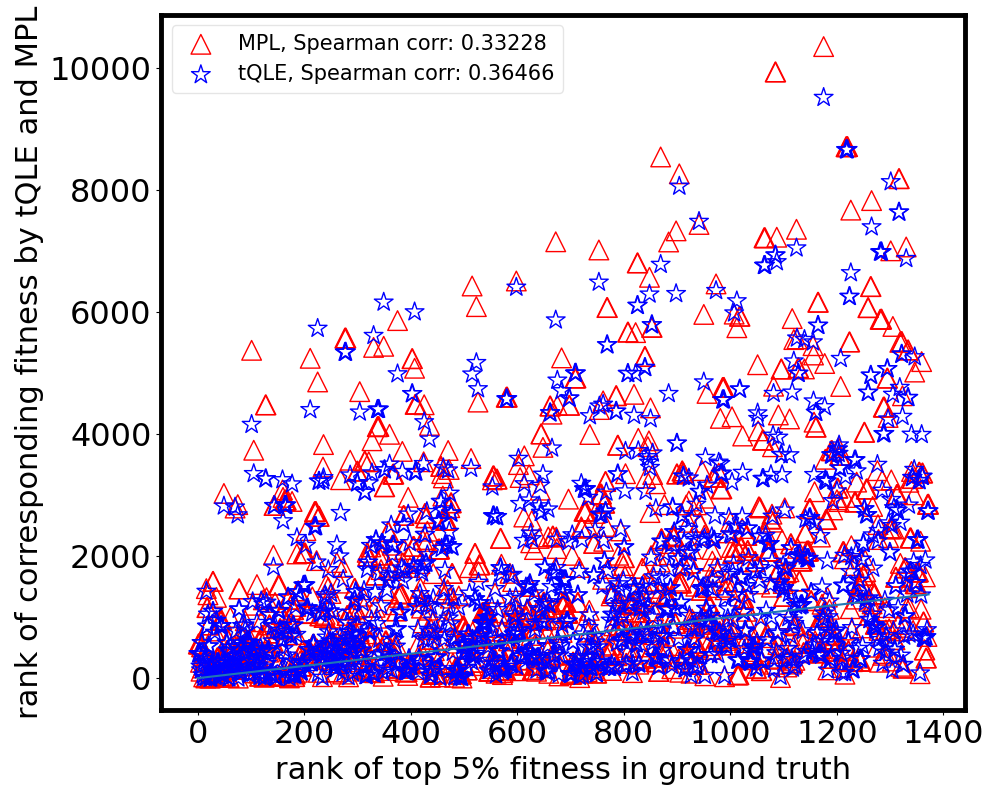}
\includegraphics[width=.3\linewidth]{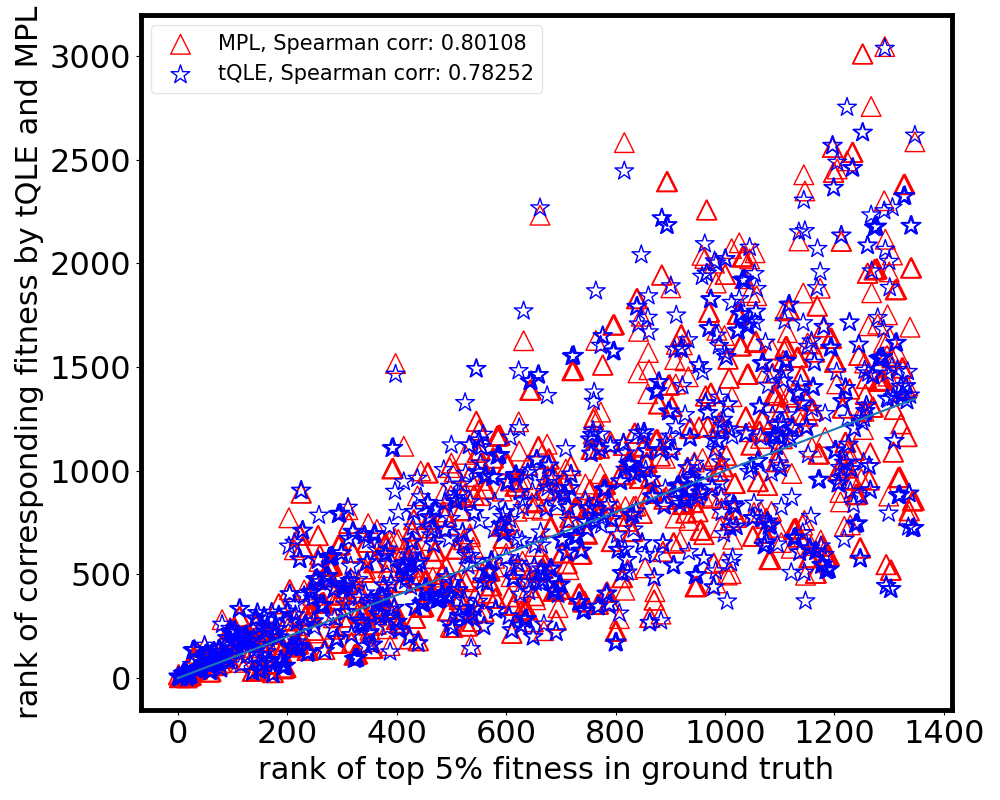}
\caption{
Scatter plot of fitness rank of top 5\%
most fit sequences
as inferred using MPL and tQLE against
ground truth.
As in Fig.~\ref{fig:scatter_plts_for_estimated_additive_fitness} ground truth fitness is due to additive fitness alone, and tQLE is used
without taking epistatic fitness into account.
Each panel displays the inferred rank (y-axis) of the top 5\% ground-truth fitness sequences (x-axis) for MPL (red triangles) and tQLE (blue stars). 
Mutation ($\mu$) and additive fitness ($\sigma(f_i)$) parameters the same as
in
Figs~\ref{fig:scatter_plts_for_fi} and~\ref{fig:scatter_plts_for_estimated_additive_fitness},
other parameter values as indicated in Table~\ref{tab:parameter-values}.
As additive effects strengthen, both methods achieve improved rank accuracy. MPL slightly outperforms tQLE under stronger selection, while the latter exhibits greater stability under lower mutations. 
}
\label{fig:scatter_plts_for_ranks_of_additive_fitness}
\end{figure}

In tasks such as protein design, identifying the highest fitness sequences is of particular interest.
Thus, we examined how well each method recovers the rank order in fitness among the top 5\% of ground-truth sequences (Fig.~\ref{fig:scatter_plts_for_ranks_of_additive_fitness}).
When the fitness effects of mutations are smaller ($\sigma(f_i)=0.01$), both methods show more substantial fluctuations in the inferred ranks of the top sequences, which are not strongly affected by changes in the mutation rate.
When the fitness effects of individual mutations are large ($\sigma(f_i)=0.05$), both tQLE and MPL obtain a strong correspondence between the inferred and true top-ranked sequences.

\subsection{Epistatic fitness landscapes}
With pairwise epistasis, the fitness of an individual genome is no longer determined solely by additive effects $f_i$, but also by interactions between pairs of loci. This means the fitness of a genotype depends not only on the sum of individual allele effects but also on how alleles at different loci interact, which can enhance or diminish overall fitness. Inference of fitness parameters in such models requires accounting for both the additive effects of alleles and the pairwise interactions between them. Therefore, inferring the fitness parameters in such models accurately is crucial for understanding complex evolutionary processes. Despite the added complexity, the inclusion of pairwise epistasis can provide a more realistic representation of evolutionary dynamics in many biological systems.

We next assess performance when fitness includes weak pairwise interactions. Here we fix a small epistatic dispersion, $\sigma(f_{ij})$, and sweep through different levels of additive dispersion, $\sigma(f_i)$, to vary the relative strength of epistatic and additive contributions to fitness while holding other evolutionary parameters constant (Table~\ref{tab:parameter-values}). Unless noted, our results aggregate $30$ independent realizations to separate overall trends from stochasticity between replicate simulations.

\begin{figure}[h!]
\centering
\includegraphics[width=.3\linewidth]{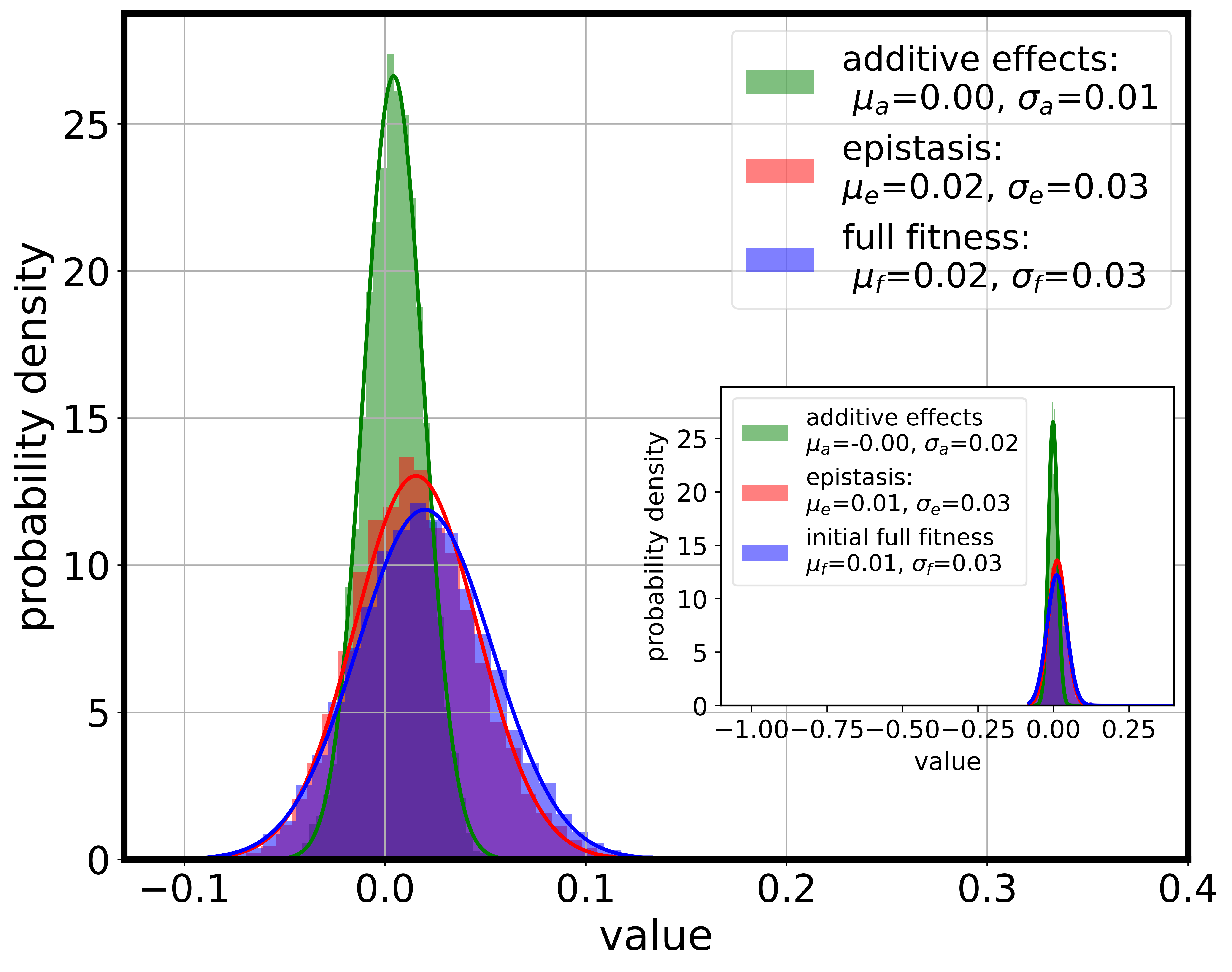}
\includegraphics[width=.3\linewidth]{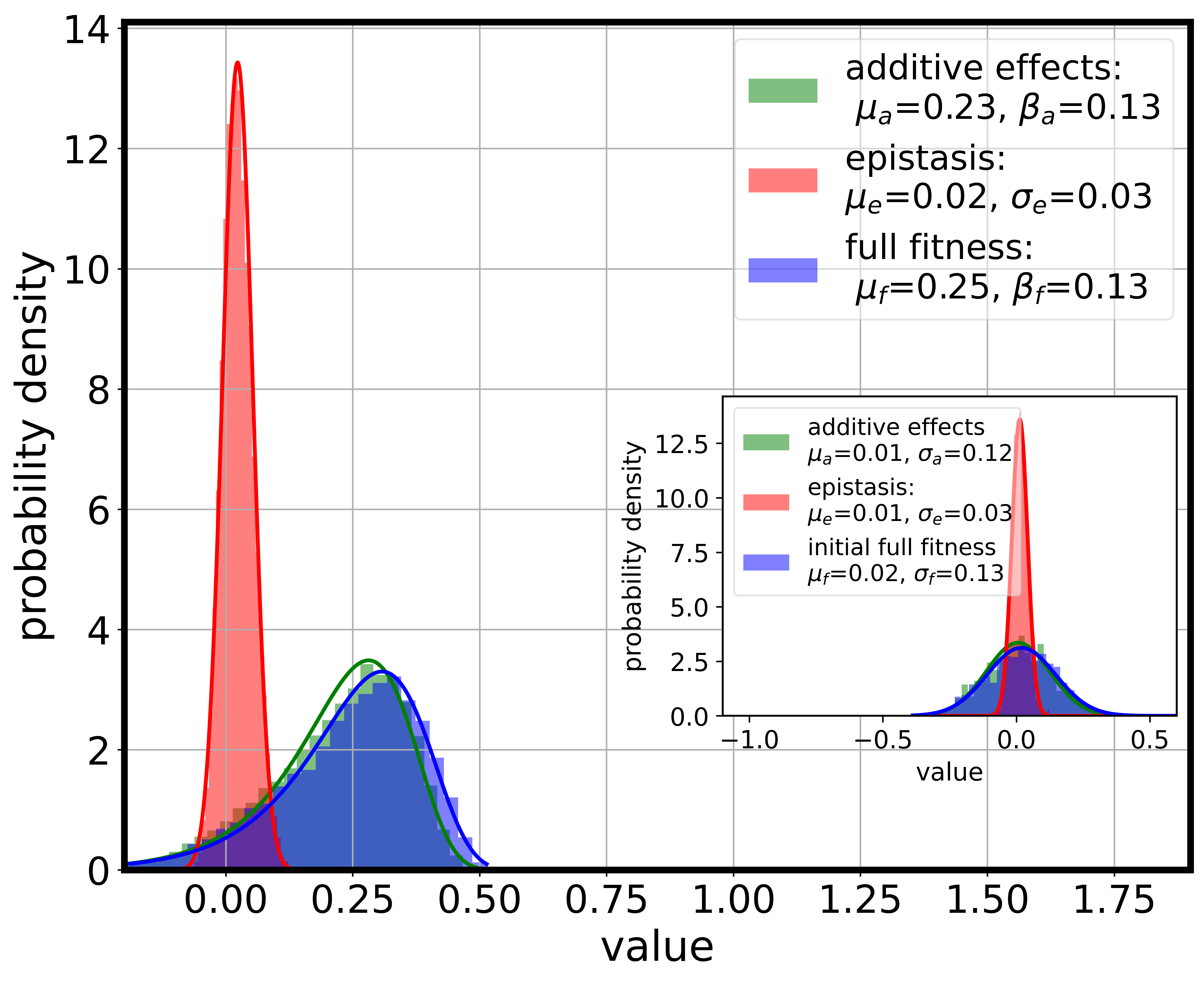}
\includegraphics[width=.3\linewidth]{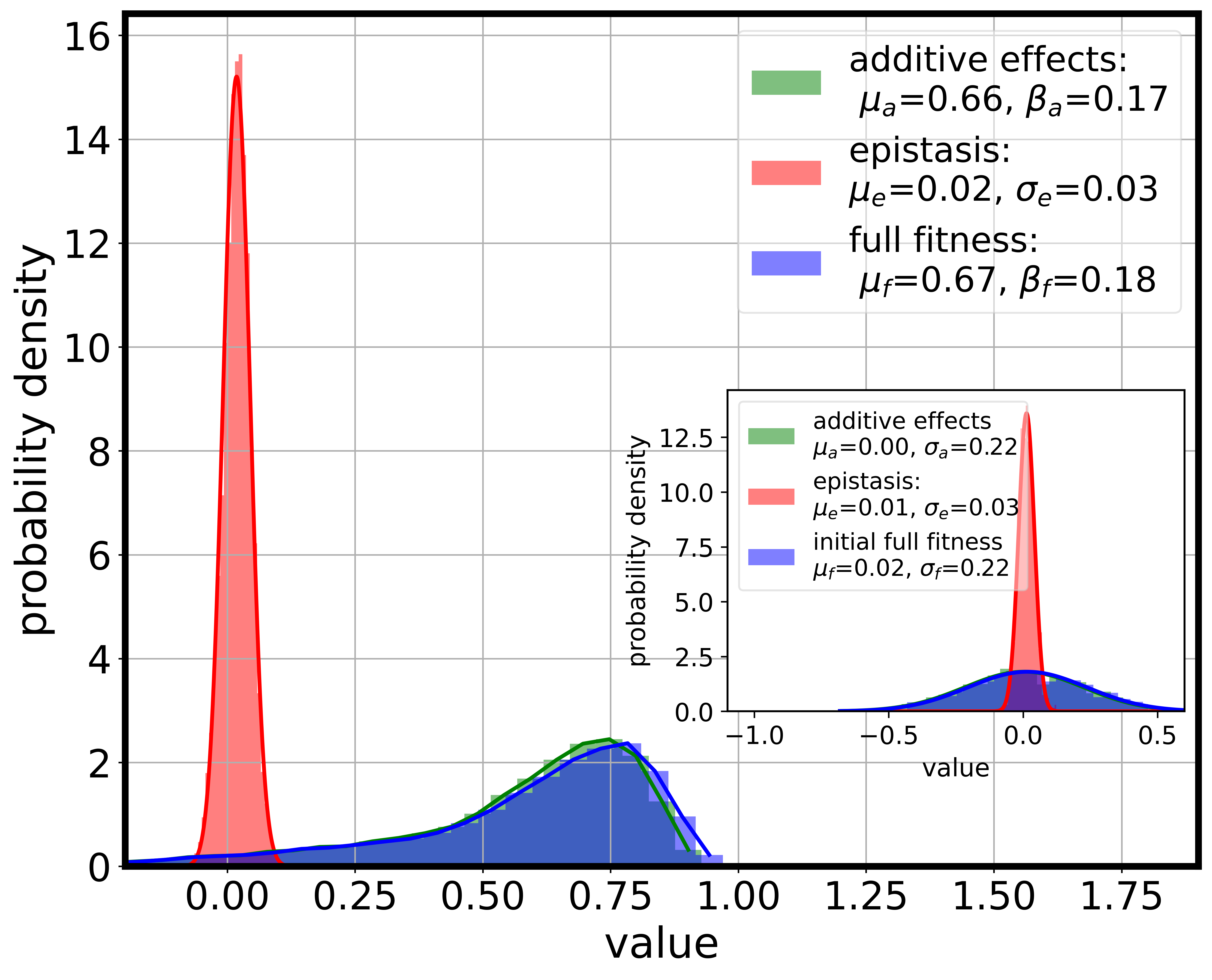}
\caption{Distributional decomposition of reconstructed fitness into additive and epistatic components under increasing variance of additive effects. Each panel shows overlaid histograms and Gaussian fits for the full fitness (magenta), additive effects (blue), and epistatic effects (red), across 30 realizations and epistatic fitness variability parameter $\sigma(f_{ij})=0.002$. Top: $\sigma(f_i)= 0.005$; middle: $\sigma(f_i)= 0.05$; bottom: $\sigma(f_i)= 0.1$.
Other parameter values as indicated in Table~\ref{tab:parameter-values}.
In low-variance settings, epistatic effects dominate fitness variation. As $\sigma(f_i)$ increases, additive effects become the primary contributor to fitness variability, reflected in broader and less symmetric full-fitness distributions.
To illustrate the deviations of final fitness distributions from Gaussians we provide numerical fits to asymmetric Gumbel (extreme value) distributions. Insets show initial fitness distributions. We note that the distributions of epistatic distributions practically do not change, in accordance with predictions in \cite{Neher2013}.}
\label{fig:histograms_for_differnt_fis}
\end{figure}

Figure \ref{fig:histograms_for_differnt_fis} shows the contributions of additive and epistatic terms to overall fitness as $\sigma(f_i)$ varies.
At low additive variance, the variability in overall fitness is dominated by epistatic contributions. As $\sigma(f_i)$ grows, the additive component becomes the primary source of variability and the full-fitness distribution broadens. This transition frames the overall inference problem: as the contribution of additive terms increases, it becomes easier to order sequences according to overall fitness, but the recovery of epistatic terms becomes more difficult.

Inserts of Fig.~\ref{fig:histograms_for_differnt_fis} show the initial fitness distribution in the population, separated into additive, epistatic and total fitness. It is clear that the distributions of epistatic fitness changes little in all three panels; epistatic fitness is not heritable when recombination rates are high, in accordance with the predictions in \cite{Neher2013}. Additive fitness on the other hand changes both as to the mean and the mode (which increase) and as to the form.
We find that a Gumbel (extreme value) distribution constitutes a resonable two-parameter fit to the final additive fitness, but note that more detailed predictions are given in \cite{Neher2013} building on the mathematical results in
\cite{Sire2006,Krug2005,Jain2007}.
Also note that the fitted curves in Fig.~\ref{fig:histograms_for_differnt_fis}
are for the extreme value distributions of a minimum, which renders their interpretation somewhat moot. Nevertheless, it is clear that the distributions  in Fig.~\ref{fig:histograms_for_differnt_fis} (middle and bottom panels) are not Gaussian, though the distribution in 
Fig.~\ref{fig:histograms_for_differnt_fis} (top panel) is close to Gaussian.

\subsubsection{Recovery of genotype-level fitness}

\begin{figure}[h!]
\centering
\includegraphics[width=.32\linewidth]{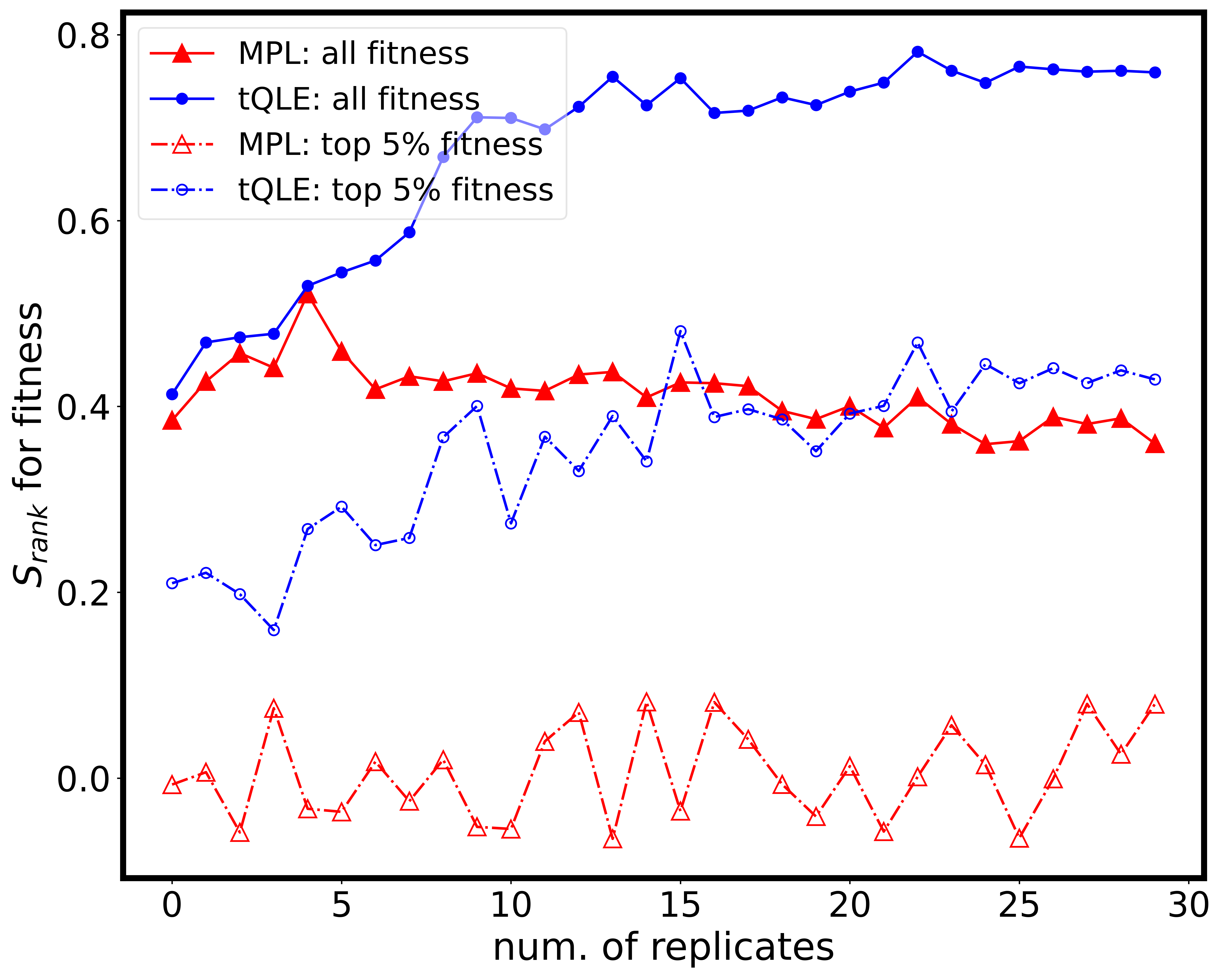}
\includegraphics[width=.32\linewidth]{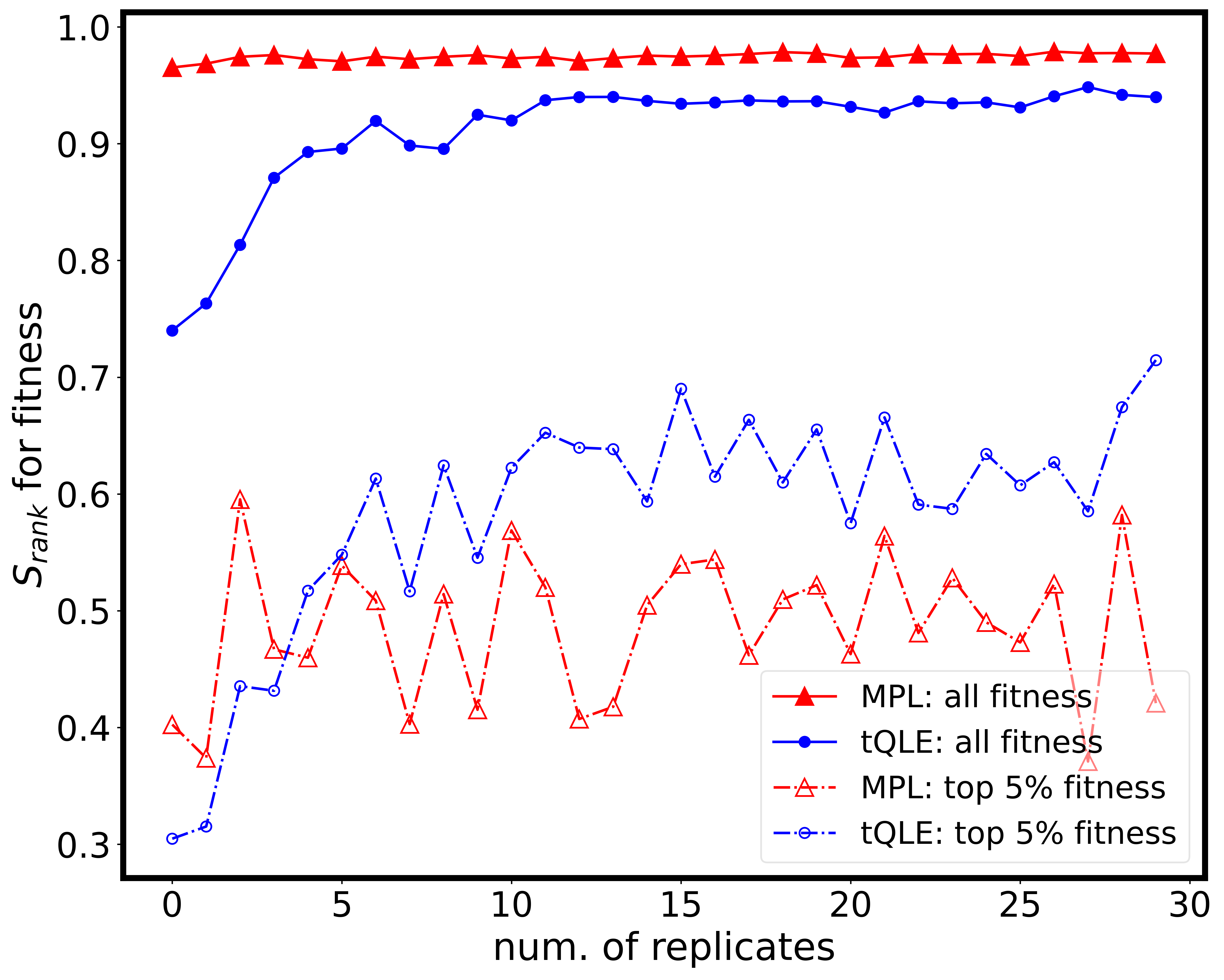}
\includegraphics[width=.32\linewidth]{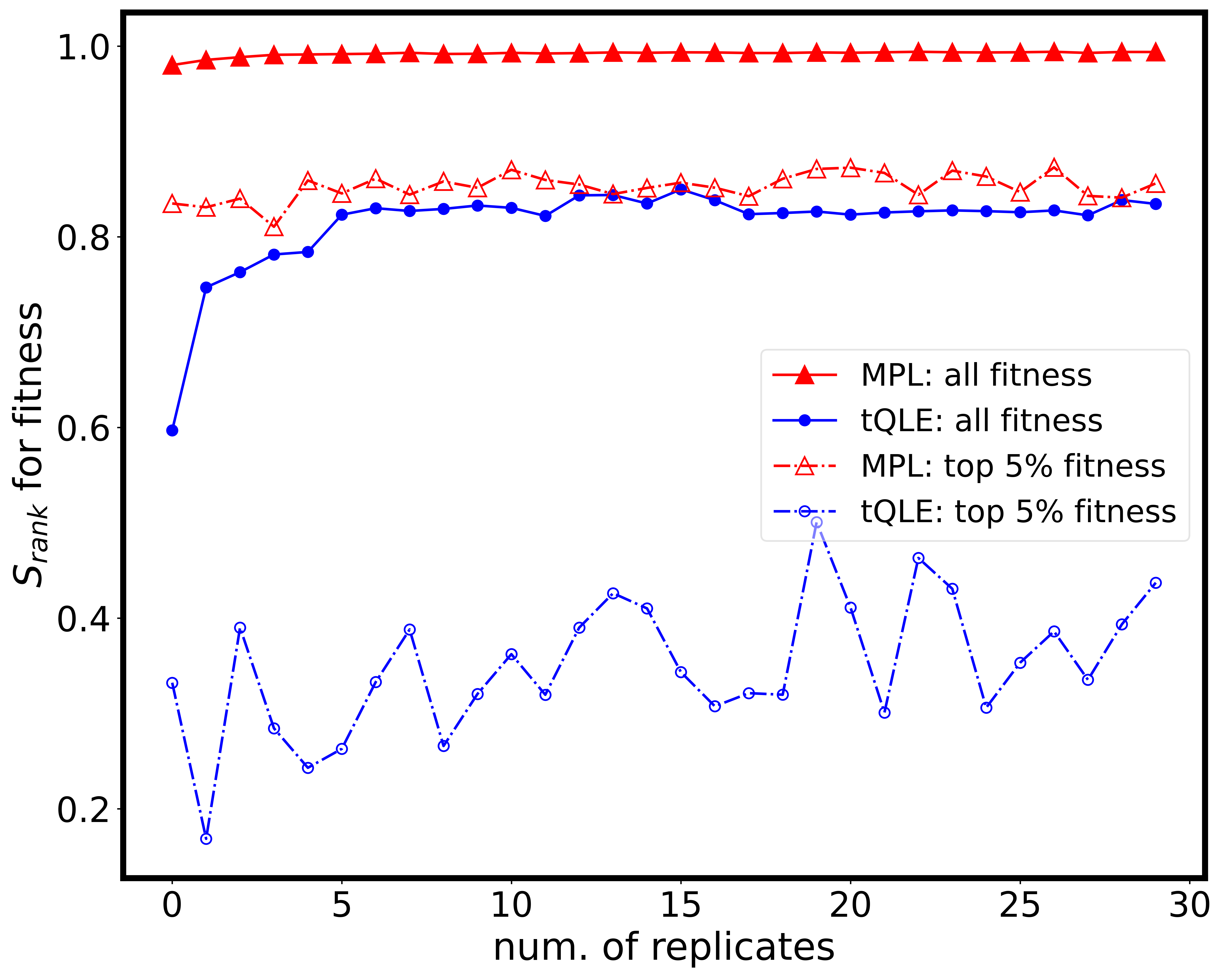}
\caption{Performance comparison between tQLE and MPL in fitness inference under different $\sigma(f_i)$. Each panel shows the Spearman correlation between the ranks of inferred fitness values compared to the ground truth for 30 independent realizations, with increasing 
additive fitness variability (parameter $\sigma(f_i)$).
Parameters are the same as
in Fig.~\ref{fig:histograms_for_differnt_fis}
and as indicated in Table~\ref{tab:parameter-values}.
Dashed lines represent the correlations of ranks of the top 5\% ground-truth fittest sequences inferred by MPL (red) and tQLE (blue), while solid lines correspond to the rank of all sequences. 
Performance of tQLE increases with increasing number of replicates while MPL is relatively insensitive to number of replicates. For lowest additive fitness (top panel), tQLE is better at inferring fitness across all sequences beyond about 5 replicates, and for the top 5\% of sequences at any number of replicates. At intermediate additive fitness (middle panel), MPL outperforms tQLE for all sequences while tQLE is marginally better for the top 5\%, at sufficiently large number of replicates. For high additive fitness (bottom panel) MPL works consistently better.
}
\label{fig:scatter_plts_for_simulation_dataset}
\end{figure}

We used Spearman correlations to quantify the overall correspondence between inferred and true fitness ranks across all genotypes and, separately, within the elite top 5\% of true fitness values (Fig.~\ref{fig:scatter_plts_for_simulation_dataset}). At low $\sigma(f_i)$, tQLE yields higher global rank correlations than MPL. As $\sigma(f_i)$ increases, MPL becomes comparatively more robust for global ordering and approaches and eventually exceeds tQLE.

Both methods are better able to recover underlying fitness values when data from multiple replicate simulations are used for inference (Fig.~\ref{fig:scatter_plts_for_simulation_dataset}), but this result is much more pronounced for tQLE. Gains in performance with increased replicates were most substantial for fitness landscapes with greater epistatic contributions to the landscape. However, in these tests, increases in performance typically saturated after around 5 replicates were included. After this point, further increasing the number of replicates had little effect on the correlation between inferred and true fitness values.

\subsubsection{Recovery of individual fitness parameters}
\begin{figure}[h!]
\centering
\includegraphics[width=.3\linewidth]{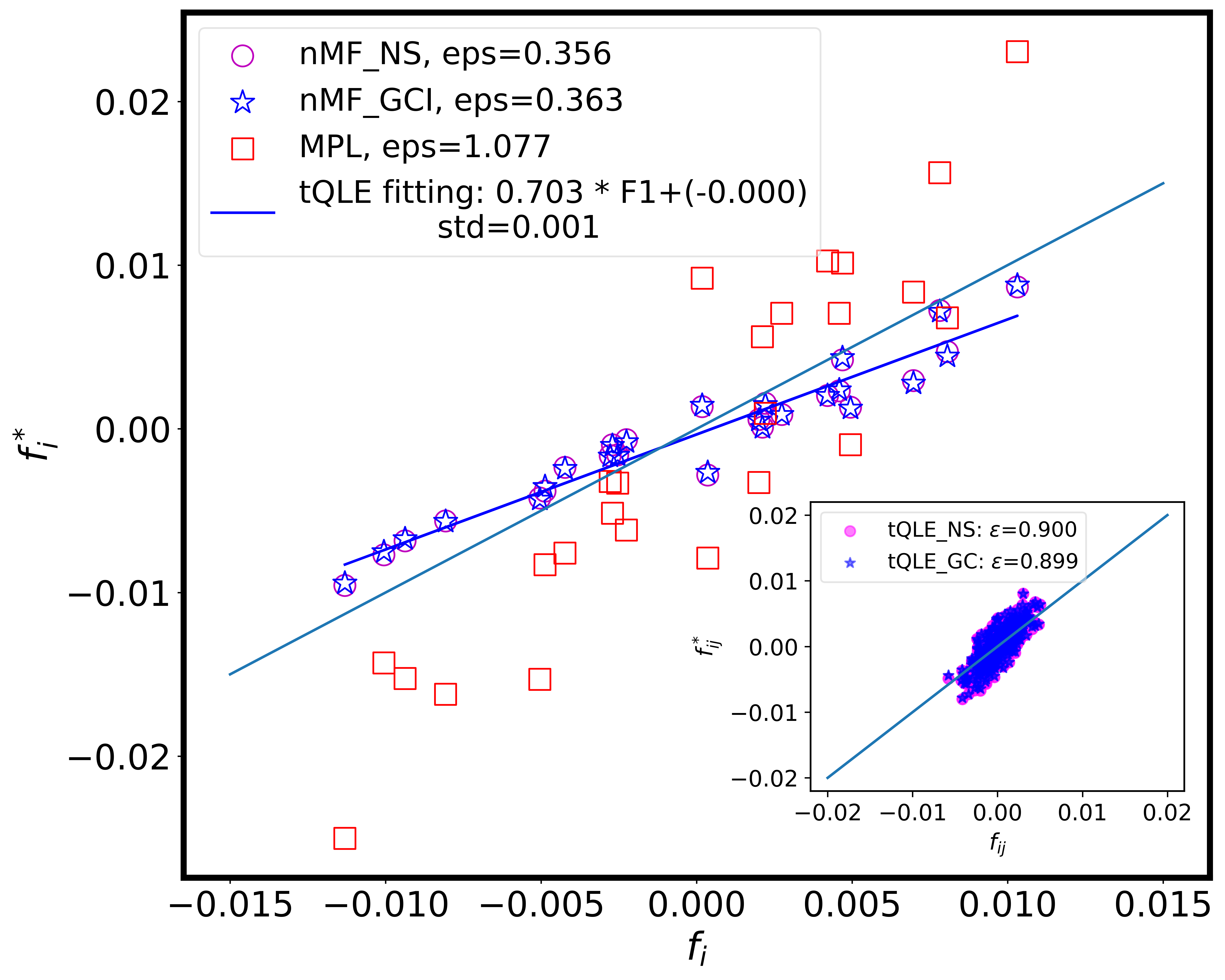}
\includegraphics[width=.3\linewidth]{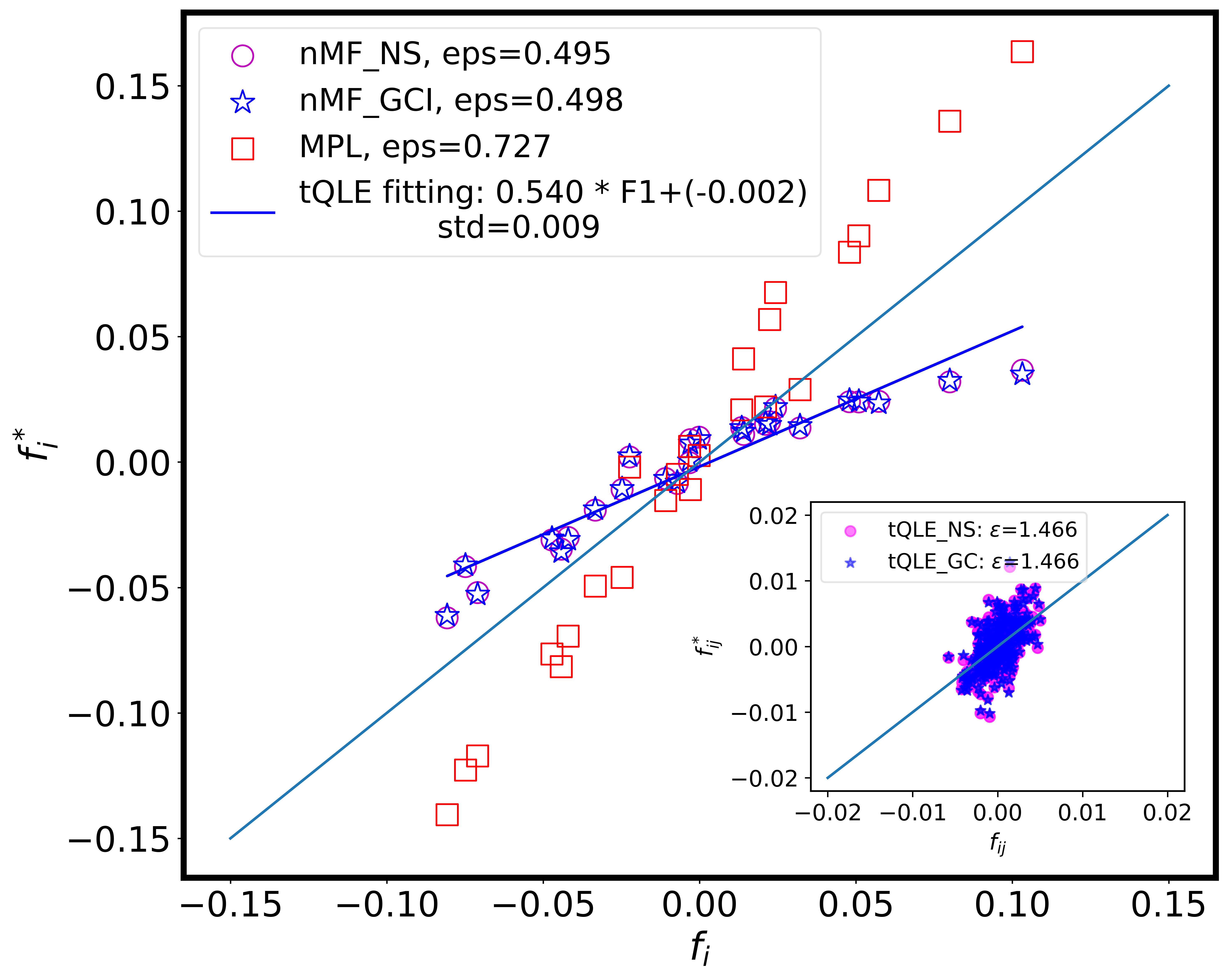}
\includegraphics[width=.3\linewidth]{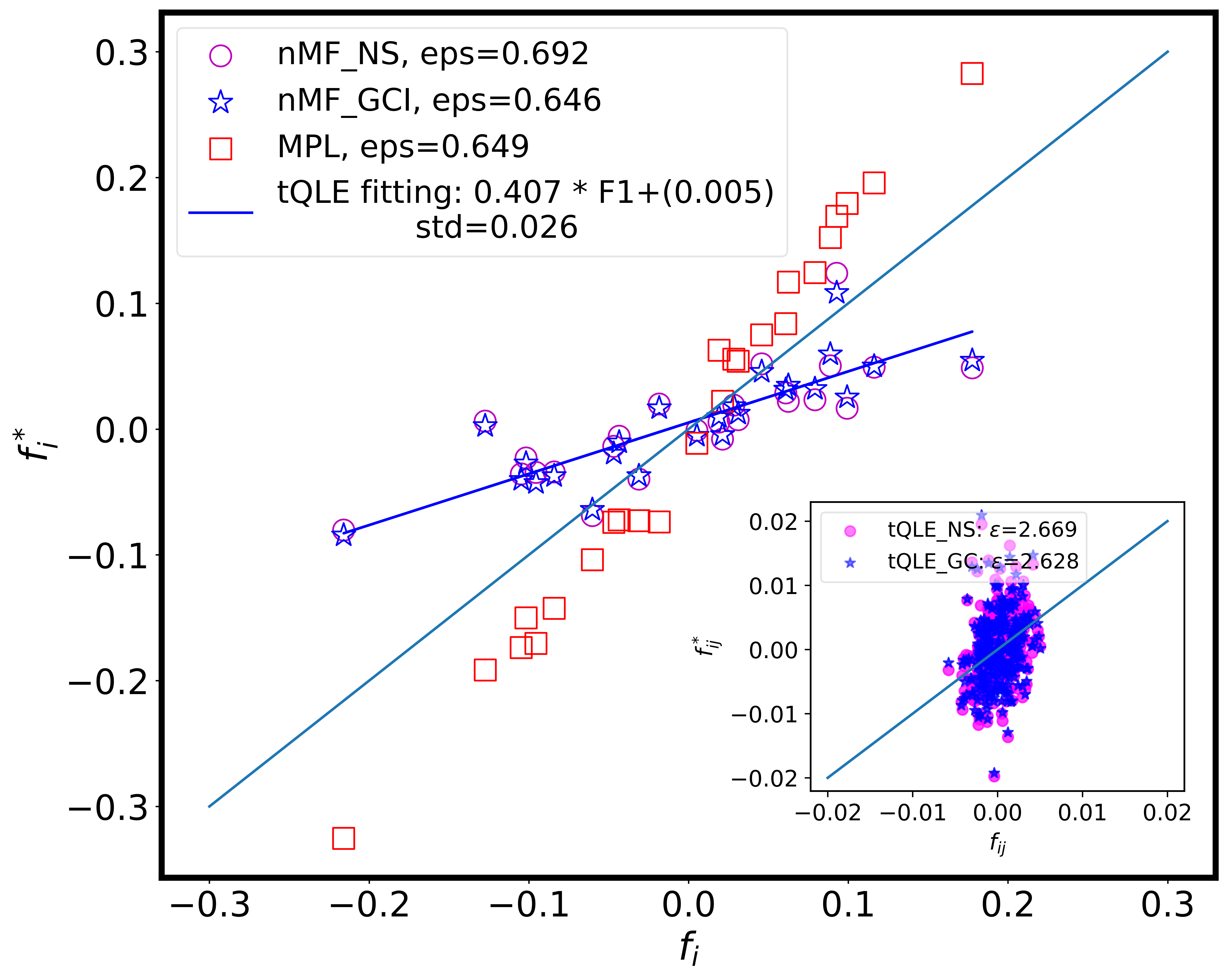}
\caption{Inferred additive effects and pairwise epistasis under increasing 
additive fitness variability parameter
$\sigma(f_i)$. 
Epistatic fitness variability parameter
$\sigma(f_{ij})$ constant; this
and other parameters the same as
in Figs.~\ref{fig:histograms_for_differnt_fis}
and~\ref{fig:scatter_plts_for_simulation_dataset},
and as indicated in Table~\ref{tab:parameter-values}.
Main plots show scatter plots of inferred additive
fitness parameters
$f_i^*$ versus the ground truth $f_i$.
The inference methods used are tQLE with reconstructed $f_{ij}^*$ by \eqref{eq:f-2-inference-GC} (blue stars),  tQLE with $f_{ij}^*$ from \eqref{eq:f-2-inference-corr} (magenta circle) and MPL (red squares).
Reconstruction errors $\varepsilon$ for each method are indicated in the figure legend. Blue line shows a linear fit of the tQLE estimates with the best-fit equation and standard deviation of the estimate indicated in the figure legend. Inserts display inference of epistatic coefficients $f_{ij}^*$ using tQLE. Magenta and blue points indicate \eqref{eq:f-2-inference-corr} and \eqref{eq:f-2-inference-GC} respectively. As selection strength increases from top to bottom, both marginal and pairwise inference performance degrade for tQLE method while MPL is getting better. 
}
\label{fig:scatter_plts_F1_F2}
\end{figure}

In addition to genotype-level fitness estimates, we also assessed the ability of tQLE and MPL to recover underlying additive and epistatic contributions to fitness from simulated evolutionary trajectories. 
Figure~\ref{fig:scatter_plts_F1_F2} shows that both approaches obtain a good linear correlation between the true and inferred additive contributions to fitness $f_i$.
Interestingly, for tQLE, recovery of additive fitness parameters becomes more challenging as the variance of the additive components $\sigma(f_i)$ increases.
As $\sigma(f_i)$ becomes larger, the slope relating the inferred and true additive parameters for tQLE becomes flatter, indicating that tQLE underestimates the magnitude of the most positive or negative $f_i$.
MPL displays the opposite trend, where estimates of additive fitness slightly improve as the additive contributions to the fitness landscape become larger.

In the tests performed here, recovery of pairwise epistatic interactions can only be assessed for tQLE, as the model for MPL assumed a purely additive fitness landscape.
Here, we estimated epistatic interactions in tQLE using both naive Mean Field (nMF) with Gaussian correction (nMF\_NS, \eqref{eq:f-2-inference-GC}) and from correlations directly (nMF\_GC, \eqref{eq:f-2-inference-corr}; see Supporting Information).
Consistent with expectations, the inference of pairwise epistasis using tQLE is most accurate at low $\sigma(f_i)$, where epistatic terms make the largest contribution to the fitness landscape. As additive variance increases, it becomes progressively more difficult for tQLE to accurately infer underlying epistatic interactions.

Consistent with these component-level trends, Fig.~\ref{fig:scatter_plts_for_fitness} shows that tQLE fitness estimates closely match the true values when additive fitness variance is low. MPL catches up and ultimately attains the best global fitness correlations at the highest tested value of $\sigma(f_i)$. However, tQLE remains a better relative predictor of fitness ranks among the highest fitness (top 5\%) sequences over a range of $\sigma(f_i)$, highlighting the potential importance of epistatic interactions in sorting the highest fitness sequences (Fig.~\ref{fig:scatter_plts_for_top_ranks_fitness}).

\section{Discussion}
\label{sec:discussion}
We have in this work considered the possibility of inferring fitness from population-wide whole-genome data so abundant that it can be stratified in the time domain. As discussed above, such data already became available in well-defined geographic domains such as the United Kingdom (except Northern Ireland) during the COVID19 pandemic. They can be expected to become available again in future pandemics, and likely also for many other pathogens in outbreaks that do not reach the pandemic level.
With the advances in ancient genomes~\cite{Paabo, Reich}, similar types of analysis may also become possible on human data, if and when the genetic make-up of a population can be consistently and continuously monitored over many generations with wide coverage.  

The core idea of the work is that the classical models of population genetics, which were invented to predict the properties and fate of biological populations evolving according to known laws with known parameters, can be turned around to inference schemes where parameters are deduced from data. 
The challenges one faces in this task range from the choice of laws considered (the problem of 'unknown unknowns'), mathematical formulations of the laws suitable for inference, the availability of data, and the proliferation of model details that could be taken into account.
We have here focused on the first two aspects and considered evolution driven by selection (fitness maximization), mutations, recombination (sex) and genetic drift (finite-size effects). All populations are taken to be well-mixed, \textit{i.e.}, geographic distance (island models) are not considered.
Also, mutation and recombination are modeled in a simplified manner using one overall uniform parameter for each.

In this setting, we have considered the possibility of inferring fitness in two ways, which both go back to pioneering contributions by Motoo Kimura in the 1950s and 1960s, respectively~\cite{Kimura1956} (see also \cite{Kimura1964}) and \cite{Kimura1965}. In marginal path likelihood (MPL) we assume that populations are characterized by probability distributions over allele frequencies which obey Kimura's diffusion equation, and the inference task is to estimate the parameters in the resulting diffusion process from time-series data. These parameters are directly related to evolutionary fitness parameters through the inference formula  
\eqref{eq:f-1-inference-MPL}.
In transient quasi-linkage equilibrium, we on the other hand assume that the probability distributions over genomes are of the Gibbs-Boltzmann type characterized by linear and quadratic terms, and not only allele frequencies.
The coefficients of these Gibbs-Boltzmann distributions are more or less indirectly related to evolutionary parameters through different approximations previously developed in the theory of the quasi-equilibrium phase \cite{NeherShraiman2011,Zeng-Mauri-2021}; we have here only used the original and simplest version of those.

One aspect that we have emphasized throughout is that inference quality (performance of an algorithm) depends on the evaluation criteria. For in silico testing, the most immediate criterion is how well one can recover model parameters, which can be visualized in a scatter plot (as we have done above). However, often one is not much interested in the numerical value of some parameter, which can also depend on regularization hyperparameters in a particular inference scheme. Rather, one is often interested in how sequences (individuals) are ordered according to total fitness, since individuals of low rank (most fit individuals) have the largest chance to survive and reproduce, whatever numerical fitness value is assigned to them.  
We have therefore also investigated how well the fitness rank of a sequence can be inferred, and in particular how well the fitness rank of the most fit individuals (sequences of lowest fitness rank) can be inferred. 

The overall conclusions of our work can be explained intuitively.
First, purely additive fitness models are generally well
captured by both methods, except when fitness effects are small and mutation rates are high. This can be attributed to lack of signal: strong mutations overpower fitness, which then cannot be recovered. Second, as MPL does not include epistatic terms, it is outperformed by tQLE when epistatic fitness is comparable to or dominates over additive fitness. Third, in the other direction, tQLE effectively assumes an infinite population, and is therefore dominated by MPL when genetic drift (finite size noise) is important. 
The results that could not have been deduced beforehand are that both models substantially agree in their predictions over rather large and reasonable parameter ranges, though still differing in some other ranges.
This is particularly so when the evaluation criterion is fitness rank.
We can therefore conclude that fitness inference is possible using both MPL and tQLE, and that it is largely a matter of convenience which method to use on real data.

As a side result we provide further support for a prediction made in \cite{Neher2013}, that epistatic fitness is not heritable, while additive fitness is heritable (inserts in the panels in Fig.~\ref{fig:histograms_for_differnt_fis}).Finally, we highlight that the availability of large data sets on pathogens and other organisms that can be stratified in time allows us to infer biological fitness also when the frequencies of alleles change over time, from appearing to reach (almost) fixation to disappearing again.

\begin{acknowledgments}
The work of HLZ was sponsored by the National Natural Science Foundation of China (11705097), the China Scholarship Council (202508320441), and the Natural Science Foundation of Nanjing University of Posts and Telecommunications (221101, 222134). The work of JPB~reported in this publication was supported by the National Institute of General Medical Sciences of the National Institutes of Health under Award Number R35GM138233. EA acknowledges support from the Swedish Research Council through grant 2020-04980.
\end{acknowledgments}

\appendix

\section{Estimated fitness by MPL and tQLE}
\begin{figure}[h!]
\centering
\includegraphics[width=.3\linewidth]{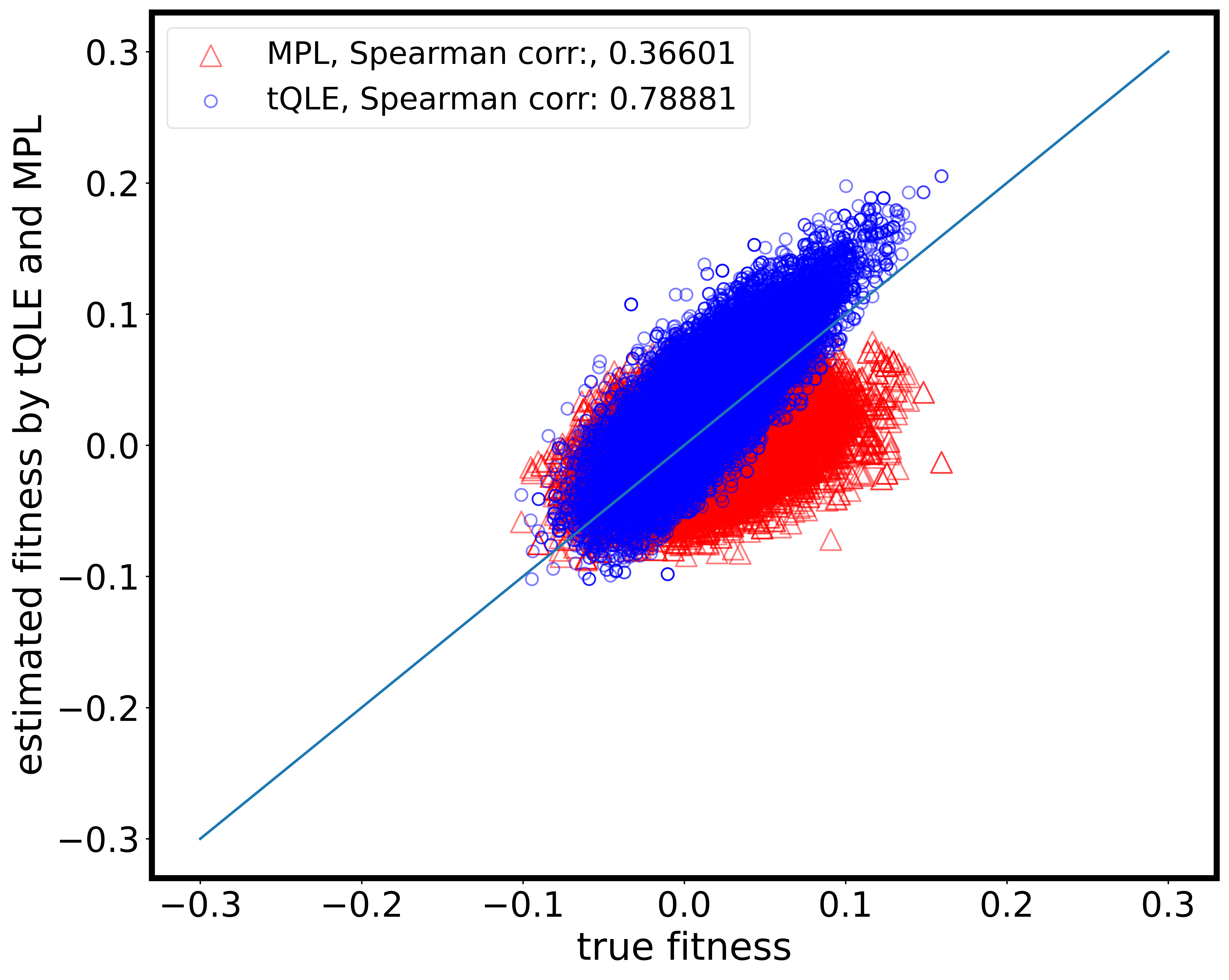}
\includegraphics[width=.3\linewidth]{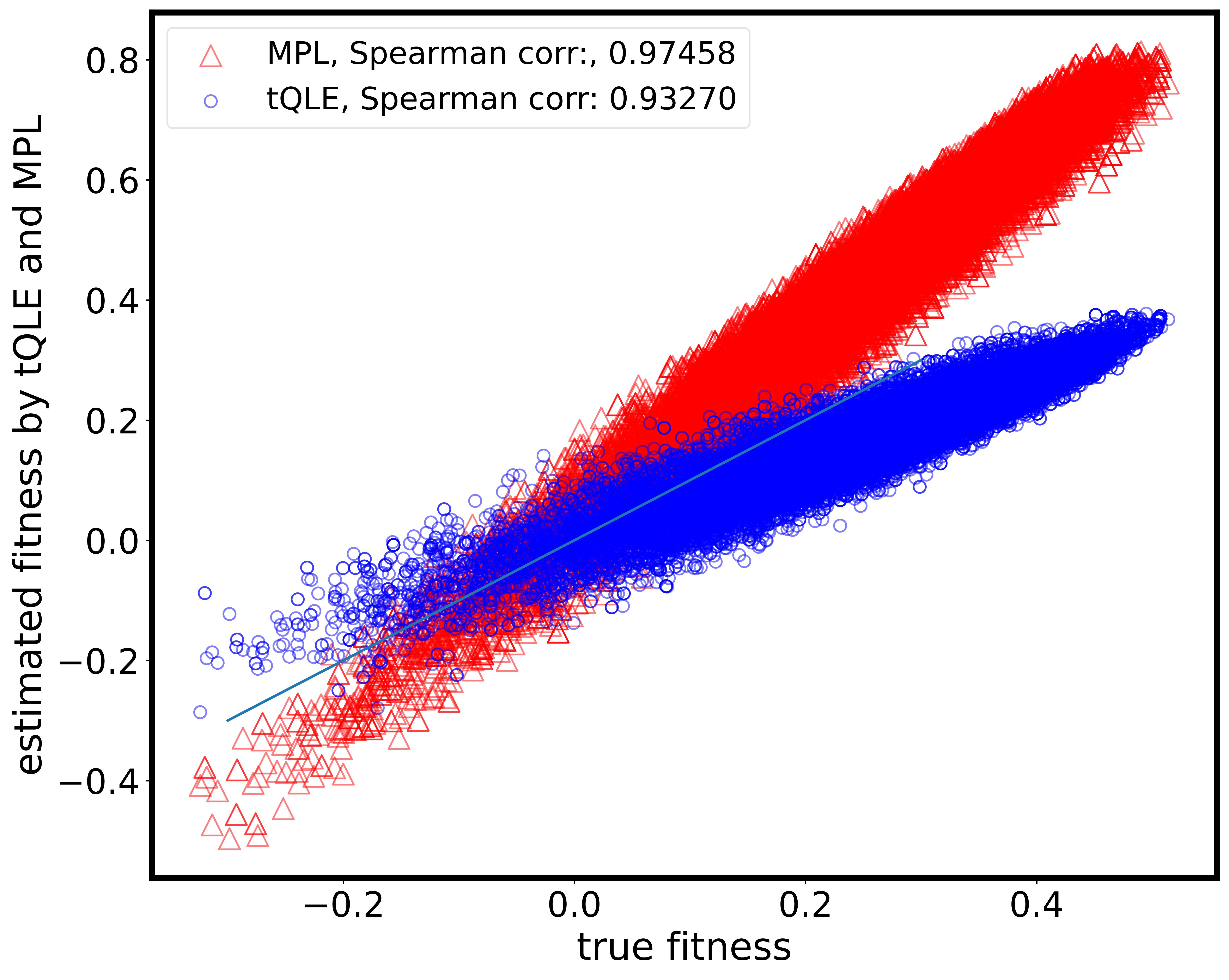}
\includegraphics[width=.3\linewidth]{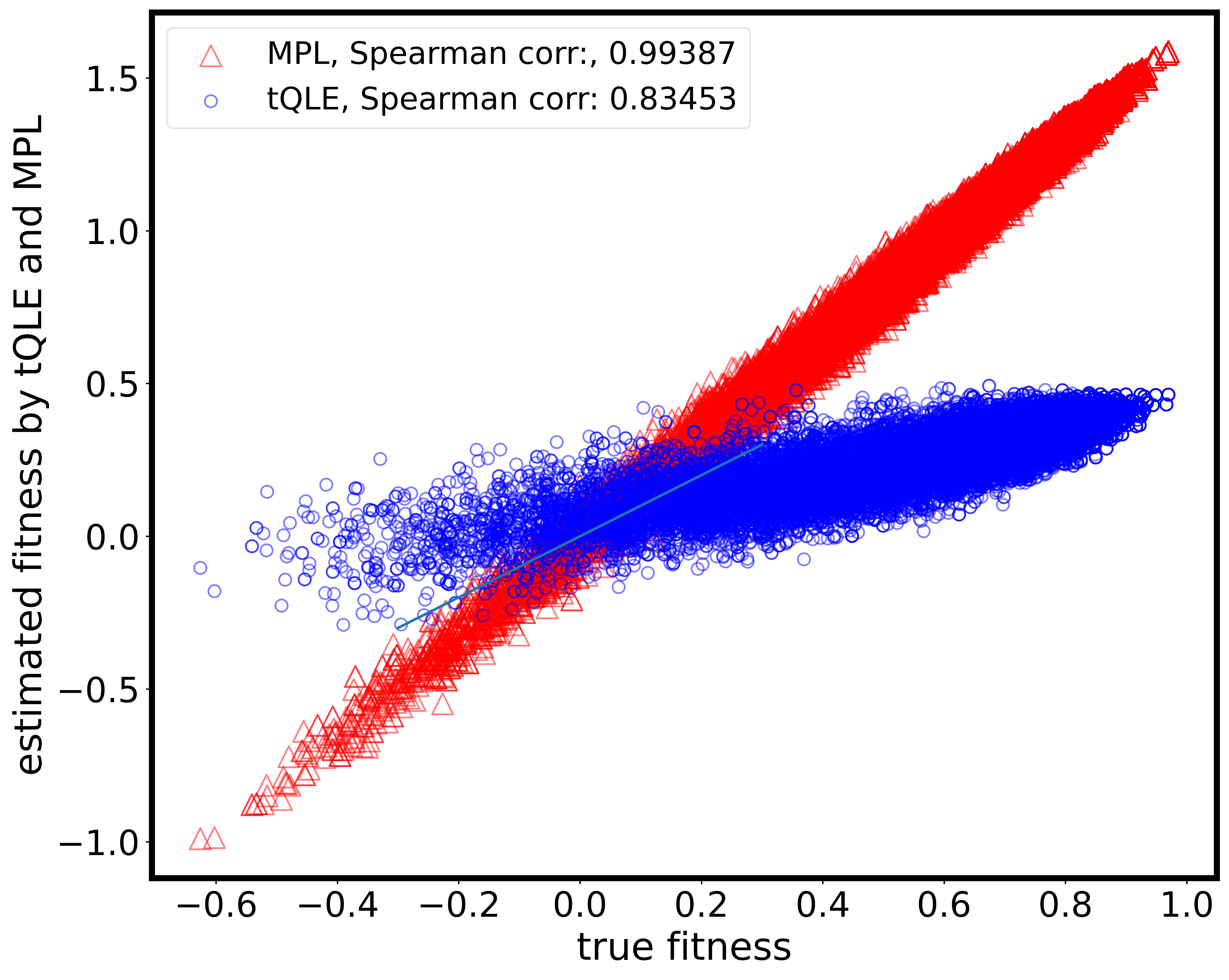}
\caption{Scatter plots comparing estimated fitness values inferred by MPL and tQLE against the ground-truth fitness.
Each panel corresponds to a different additive fitness noise level: top ($\sigma(f_i) = 0.005$), middle ($\sigma(f_i) = 0.05$), and bottom ($\sigma(f_i) = 0.1$) and a constant epistasis $\sigma(f_{ij})=0.002$. 
Other parameter values as in Table~\ref{tab:parameter-values-SI}.
Red triangles represent fitness estimates by MPL; blue circles represent estimates by tQLE. The horizontal axis shows the true fitness, and the vertical axis shows the estimated values. Spearman correlations are provided for each method. 
QLE demonstrates a tighter functional relationship (smaller scatter) while consistently underestimating the 
numerical values at low fitness (blue point clouds, top to bottom).
MPL fails at low fitness values (top panel)
while correctly capturing both functional relationship and numerical fitness values at moderate and high fitness values (middle and bottom panel).
}
\label{fig:scatter_plts_for_fitness}
\end{figure}

Using the genotypes obtained from 30 independent replicates, we conduct a detailed analysis of the estimated additive and epistatic fitness components in comparison with the ground truth. Figure \ref{fig:scatter_plts_for_fitness} compares the estimation accuracy of fitness values inferred by two methods, MPL (red triangles) and tQLE (blue circles) under varying additive effects. The parameter settings are identical to those used in the previous figures for the epistatic case. 

In the top panel ($\sigma(f_i) = 0.005$), tQLE exhibits a strong linear correlation with the ground truth (correlation = 0.95). In contrast, MPL fails to capture the full fitness, resulting in low correlation (0.39). With $\sigma(f_i) = 0.05$ (middle panel), tQLE maintains a higher correlation (0.92) compared to MPL (0.87), demonstrating more accurate full fitness inference. With $\sigma(f_i) = 0.1$ ( bottom panel), MPL achieves the highest correlation (0.91). While the performance of tQLE begins to degrade (correlation = 0.78), indicating that at higher additive effects, MPL becomes relatively more robust while tQLE loses its advantage.

\section{Inferring rank order of top-performing sequences} 
\begin{figure}[h!]
\centering
\includegraphics[width=.3\linewidth]{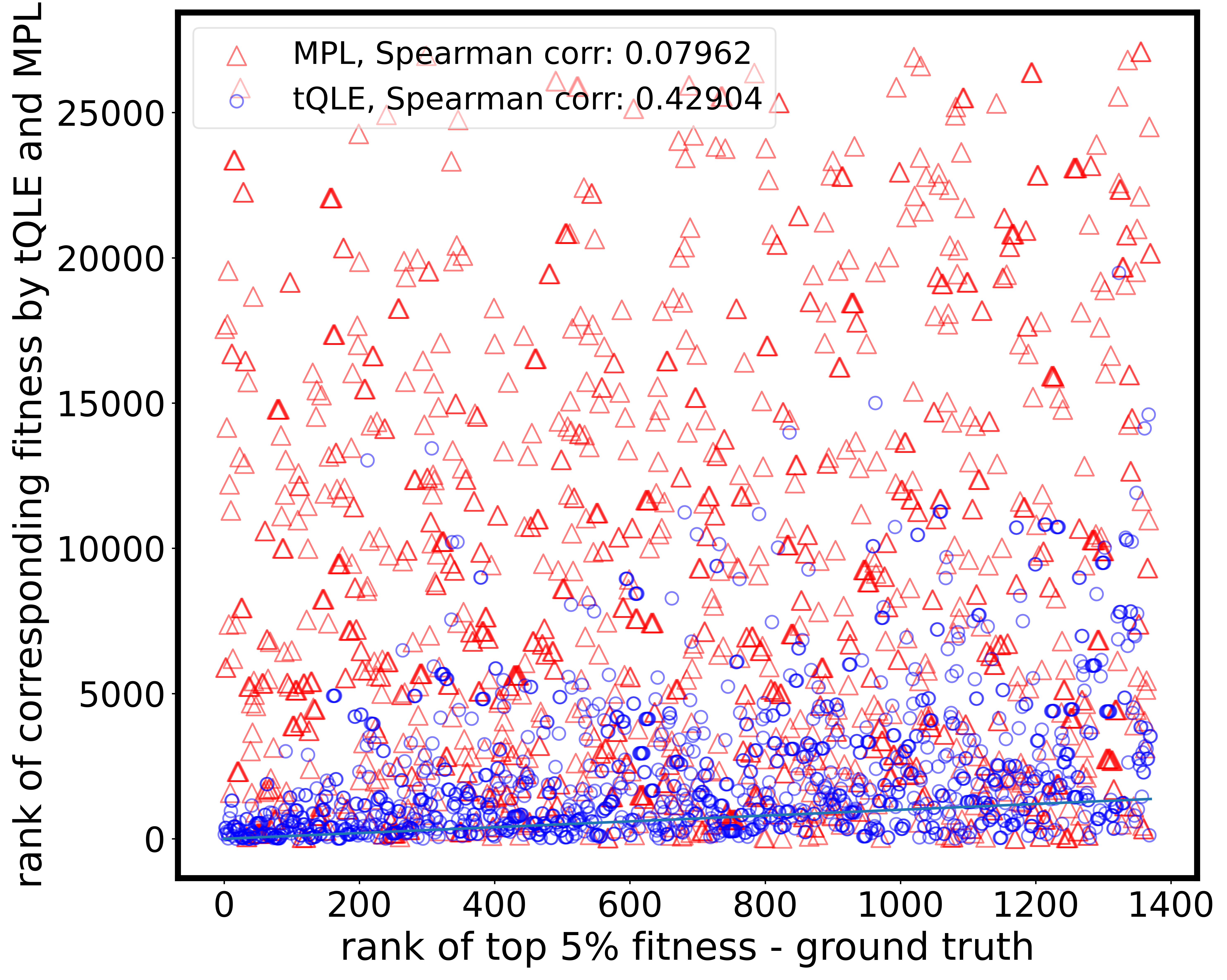}
\includegraphics[width=.3\linewidth]{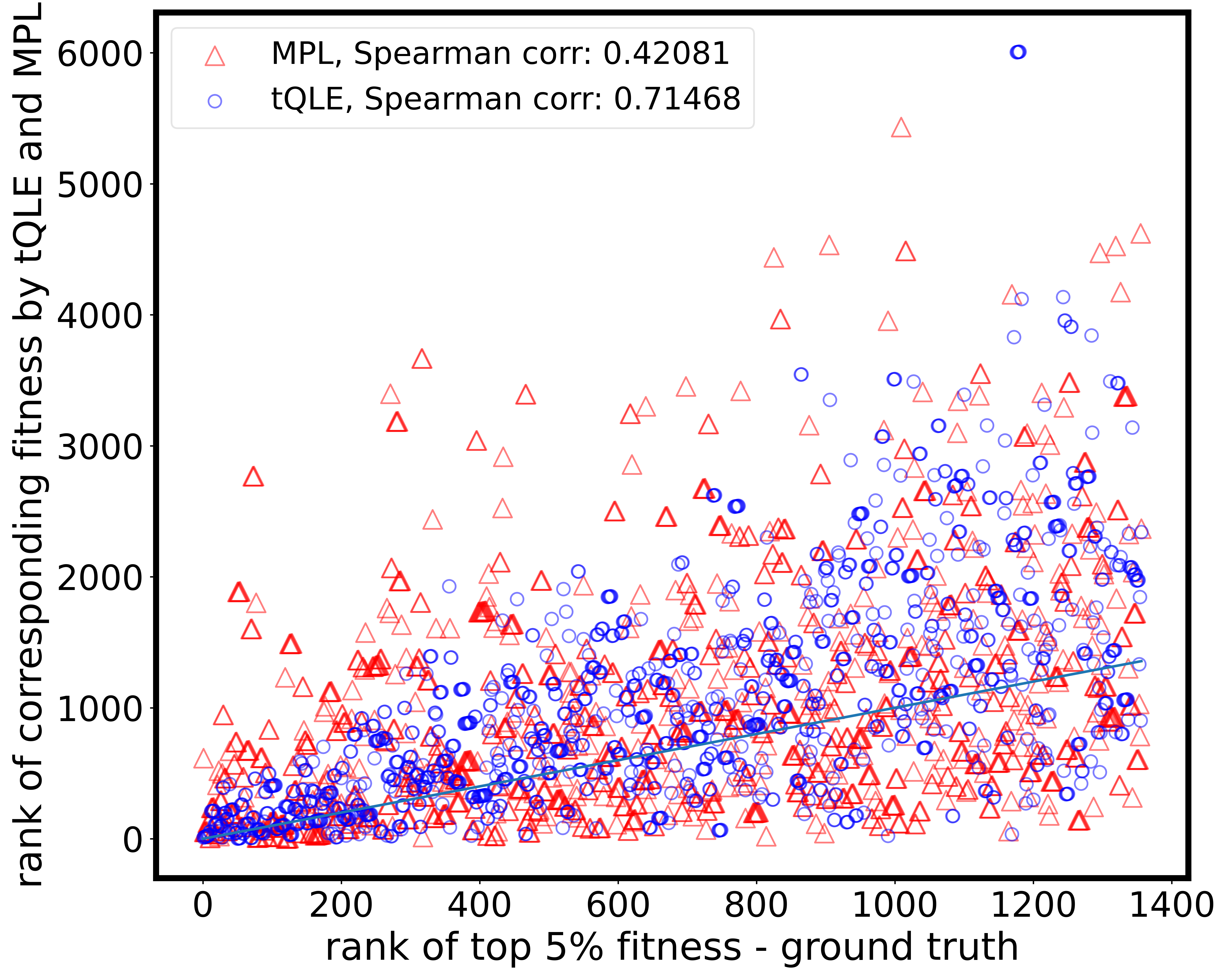}
\includegraphics[width=.3\linewidth]{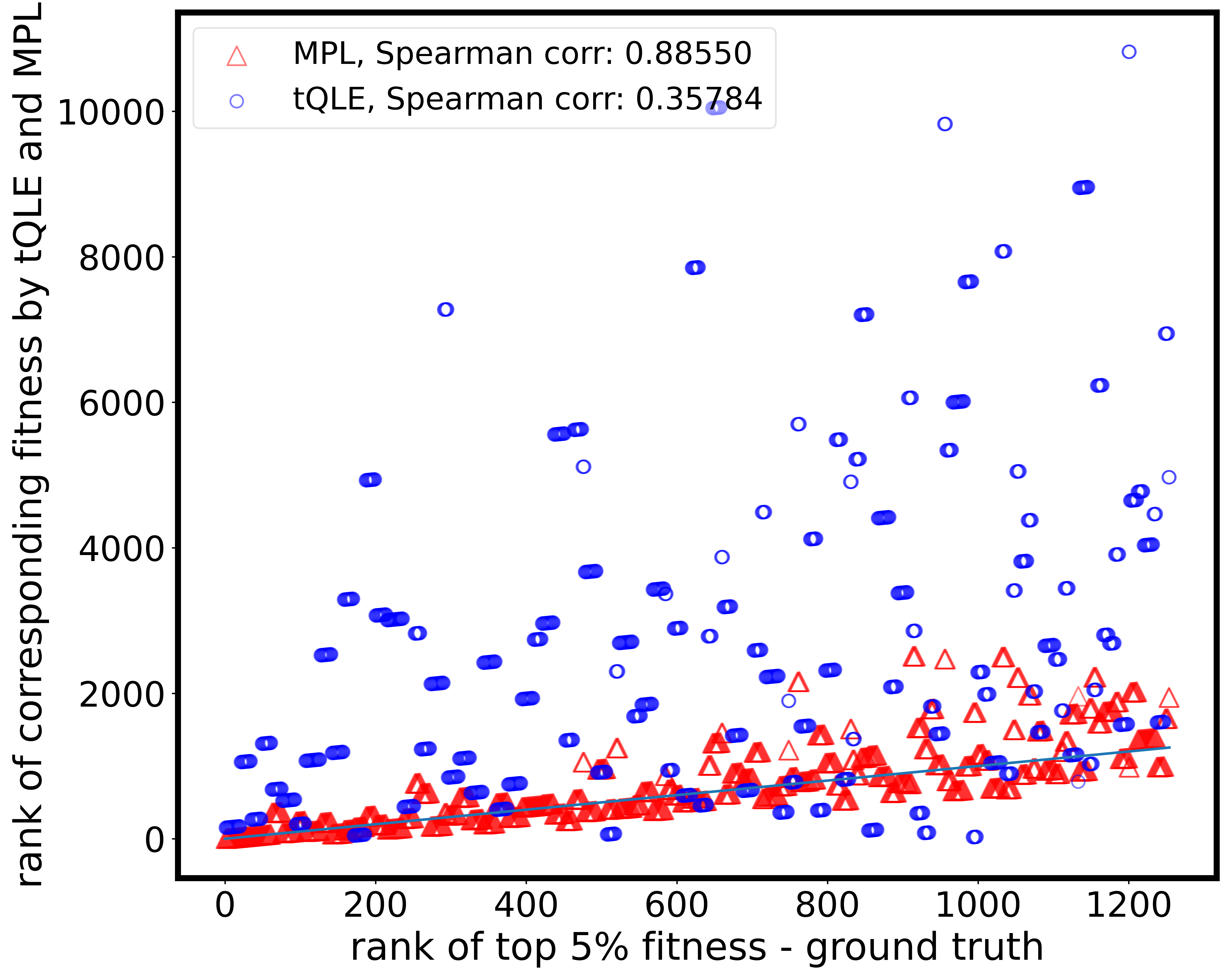}
\caption{Comparison of inferred ranks for top 5\% high-fitness sequences using MPL and tQLE under increasing additive effects. Each subplot displays the rank of inferred fitness (y-axis) versus the true rank of fitness (x-axis) for the top 5\% of genotypes, with additive fitness noise $\sigma(f_i)$ set to 0.005 (top), 0.05 (middle), and 0.1 (bottom). 
Other parameter values as in 
in Fig.~\ref{fig:scatter_plts_for_fitness}
and as indicated in
Table~\ref{tab:parameter-values-SI}.
Red triangles denote MPL results; blue circles denote tQLE results. While MPL consistently shows poor correspondence with the ground truth, tQLE achieves higher rank accuracy, particularly at lower noise levels. Correlation coefficients for each method are indicated in the legends.
}
\label{fig:scatter_plts_for_top_ranks_fitness}
\end{figure}

To further evaluate the inference quality of MPL and tQLE, we examined their ability to recover the rank order of the top-performing sequences. With the same genotypes obtained from 30 independent replicates, Fig. \ref{fig:scatter_plts_for_top_ranks_fitness}  presents scatter plots of the inferred fitness ranks against the true ranks for the top 5\% of genotypes, under three levels of additive effects: $\sigma(f_i) = 0.005$, 0.05, and 0.1. Across all panels, red triangles represent predictions from MPL, while blue circles correspond to tQLE estimates. The true fitness ranks are shown on the x-axis, and the corresponding inferred ranks are shown on the y-axis. Spearman correlations between inferred and true ranks are reported for both methods.

In the low-noise regime (top panel, $\sigma(f_i) = 0.005$), MPL performs poorly, exhibiting high variance and low correlation ($r = 0.16$). In contrast, tQLE provides a markedly better fit to the ground truth ($r = 0.68$), with rank predictions concentrated near the diagonal. As the $\sigma(f_i)$  increases (middle panel, 0.05), tQLE maintains its superior ranking performance ($r = 0.63$), whereas MPL remains largely ineffective ($r = 0.10$). With even larger $\sigma(f_i)=0.1$ (bottom panel), the performance of tQLE begins to degrade ($r = 0.46$), but it still consistently outperforms MPL ($r = 0.13$).

\begin{figure}[h!]
\centering
\includegraphics[width=.3\linewidth]{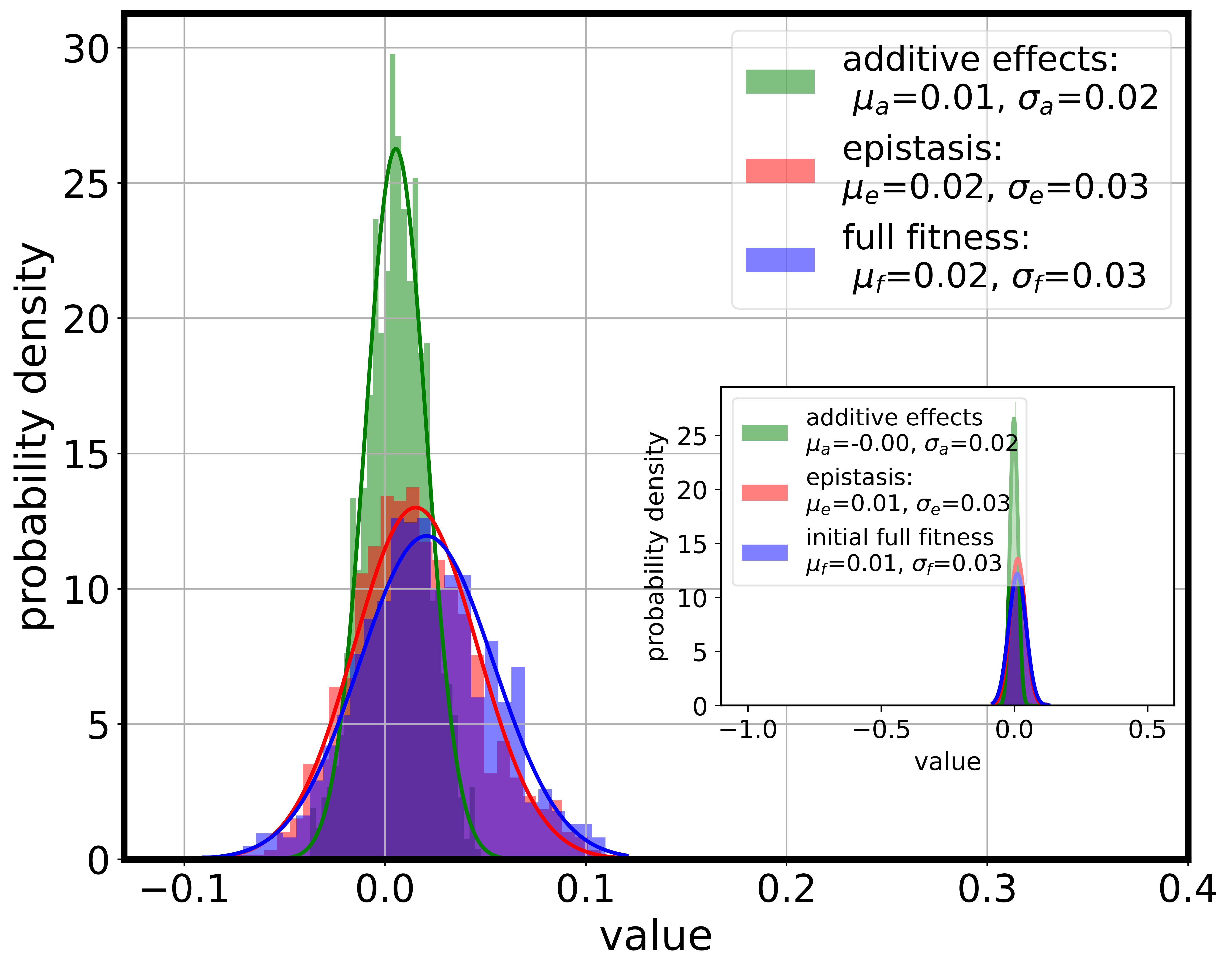}
\includegraphics[width=.3\linewidth]{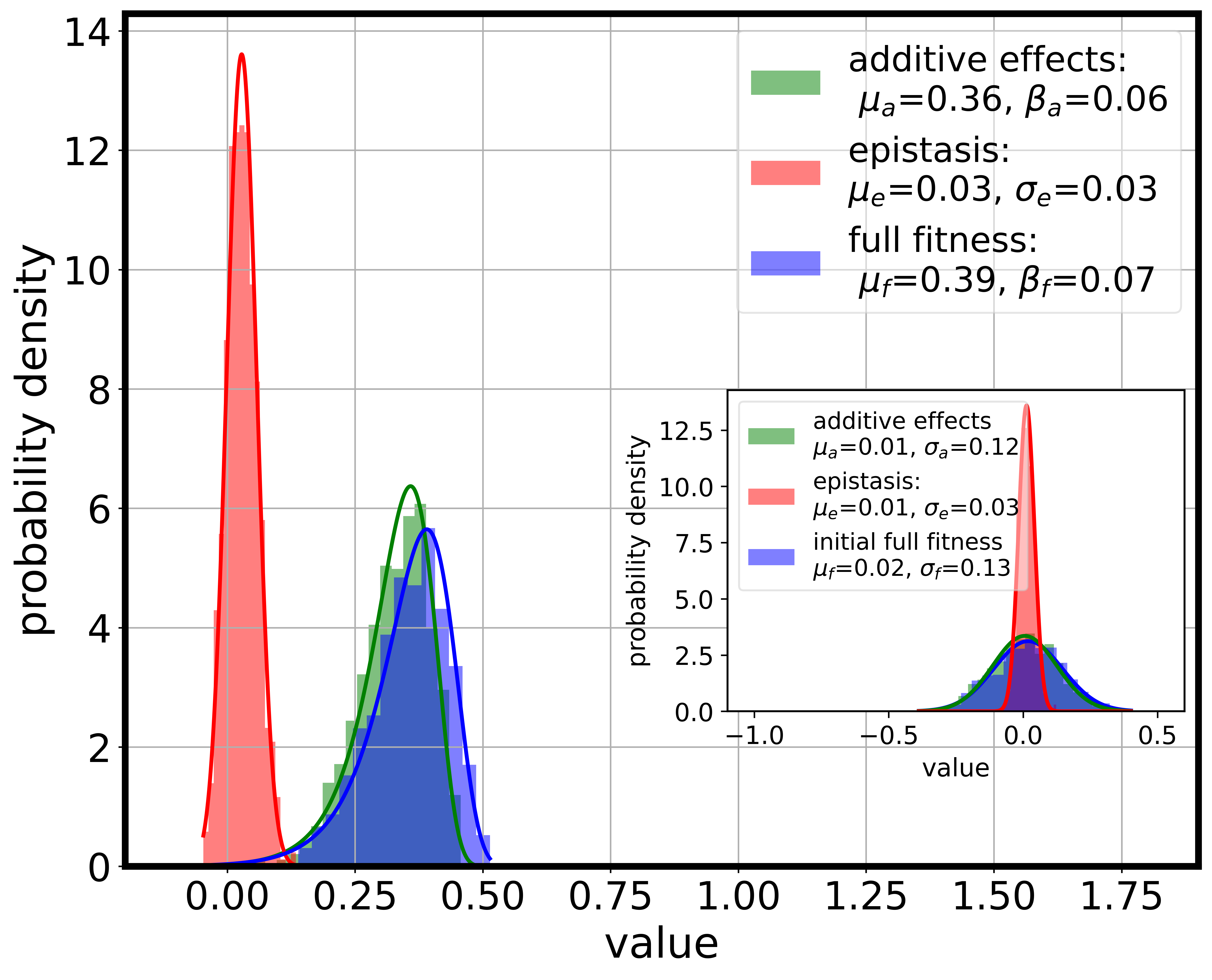}
\includegraphics[width=.3\linewidth]{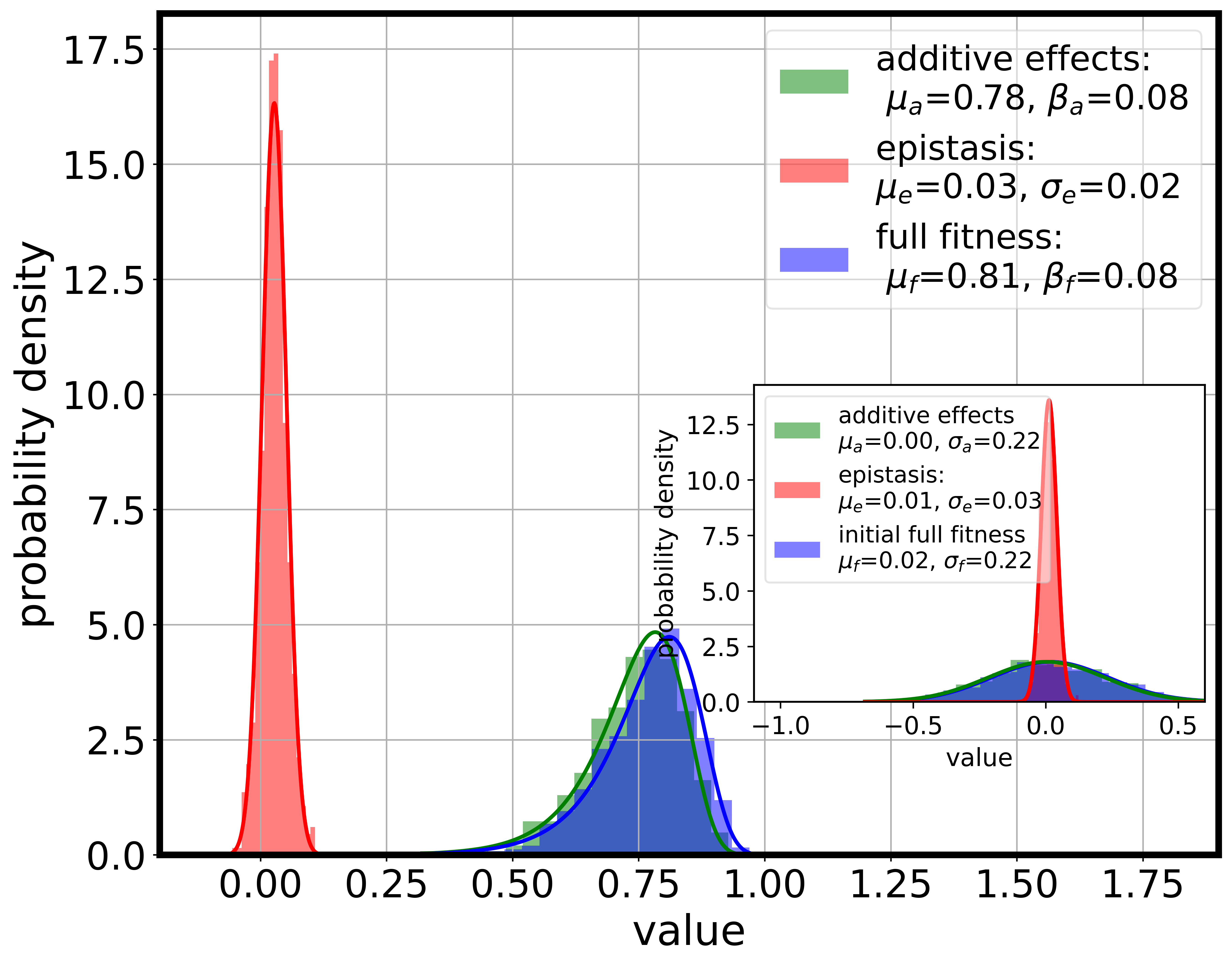}
\caption{Distributions over the initial fitness (inner panel) and one evolutionary fitness with ground truth fitness parameters.}
\label{fig:Dist_fitness_single_generation}
\end{figure}

\begin{table*}[h!]
\caption{\label{tab:parameter-values-SI}Summary of simulation parameters in simulations of evolving populations in Supplementary Information}
\begin{ruledtabular}
\centering
\begin{tabular}{l|ccccllll}
Symbol & $N$ & $N_{\mathrm{realizations}}$ & $L$ &
$T$ & $r$ & $\mu$ & $\sigma(f_i)$ & $\sigma(f_{ij})$
\\
Name & Population  & Replicates & Genome &
Time & Recombination & Mutation  & Additive  & Epistatic 
\\
   &  size &   & size &
 &  rate &  rate &  fitness &  fitness
\\
\hline
Fig~\ref{fig:scatter_plts_for_fitness},
\ref{fig:scatter_plts_for_top_ranks_fitness} &
1000,1000,1000 & 30,30,30 &
25,25,25 & 30,30,30 & 0.5,0.5,0.5 & 
0.01,0.01,0.01 & 0.005,0.05,0.1 & 0.002,0.002,0.002
\\
\end{tabular}
\end{ruledtabular}
\footnotetext{This table summarizes and completes 
the descriptions of simulation parameters given in figure caption to respective figure. Multiple numerical values indicate values in different panels, top to bottom.
In Fig~\ref{fig:scatter_plts_for_fitness}
and Fig~\ref{fig:scatter_plts_for_top_ranks_fitness}
the parameter varied is $\sigma(f_i)$,
the variability (standard deviation) of the
additive fitness parameters $F_i$ (Gaussian random numbers).
}
\end{table*}

\clearpage      
\FloatBarrier 
\bibliography{fitness-inference}

\end{document}